\documentclass[11pt,a4paper]{article}
\pdfoutput=1

\usepackage[utf8]{inputenc}
\usepackage{a4wide}
\usepackage{amsmath}
\usepackage{cite}
\usepackage{bm}
\usepackage{color}
\usepackage{braket}
\usepackage{colordvi}
\usepackage{amsfonts}
\usepackage{booktabs}
\usepackage{multirow}
\usepackage{pdflscape}
\usepackage{textcomp}
\usepackage{gensymb}
\usepackage{bigstrut}
\usepackage{array}
\usepackage{enumerate}
\usepackage[small,bf]{caption}
\usepackage{graphicx}
\usepackage{mathbbol}
\usepackage{amssymb}
\usepackage{appendix}
\usepackage{verbatim}
\usepackage{geometry}
\usepackage{makecell}
\usepackage{listings}
\usepackage{mathtools}
\usepackage{dsfont}
\usepackage{accents}
\usepackage{longtable}
\usepackage{colortbl}
\usepackage{braket}
\usepackage{enumitem}
\usepackage[colorlinks, linkcolor=black, anchorcolor=black, citecolor=green]{hyperref}

\newcommand*{\belowrulesepcolor}[1]{%
  \noalign{%
    \kern-\belowrulesep
    \begingroup
      \color{#1}%
      \hrule height\belowrulesep
    \endgroup
  }%
}
\newcommand*{\aboverulesepcolor}[1]{%
  \noalign{%
    \begingroup
      \color{#1}%
      \hrule height\aboverulesep
    \endgroup
    \kern-\aboverulesep
  }%
}
\definecolor{light-gray}{gray}{0.9}
\definecolor{lighter-gray}{gray}{0.95}

\allowdisplaybreaks
\numberwithin{equation}{section}

\begin{document}

%%%%%%%%%%%%%%%%%%%%%%%
\begin{titlepage}

\begin{center}
{\bf\LARGE Interplay and Correlations Between Quark and Lepton Observables in Modular Symmetry Models
}\\[8mm]

Gui-Jun~Ding\(^{\,a,}\)\footnote{E-mail: \texttt{dinggj@ustc.edu.cn}},
Eligio~Lisi\(^{\,b,}\)\footnote{E-mail: \texttt{eligio.lisi@ba.infn.it}},
Antonio~Marrone\(^{\,c,b,}\)\footnote{E-mail: \texttt{antonio.marrone@uniba.it}},
S.~T.~Petcov\(^{\,d,e,}\)\footnote{E-mail: \texttt{petcov@sissa.it}, Also at
Institute of Nuclear Research and Nuclear Energy,
Bulgarian Academy of Sciences, 1784 Sofia, Bulgaria.}\\
 \vspace{5mm}
\(^{a}\)\,{\it Department of Modern Physics, University of Science and Technology of China,\\
Hefei, Anhui 230026, China} \\
\vspace{2mm}
\(^{b}\)\,{\it Istituto Nazionale di Fisica Nucleare, Sezione di Bari, %\\
Via Orabona 4, 70126 Bari, Italy}\\
\vspace{2mm}
\(^{c}\)\,{\it Dipartimento Interateneo di Fisica ``Michelangelo Merlin,'' %\\
Via Amendola 173, 70126 Bari, Italy}\\
\vspace{2mm}
\(^{d}\)\,{\it INFN/SISSA, Via Bonomea 265, 34136 Trieste, Italy}\\
\vspace{2mm}
\(^{e}\)\,{\it Kavli IPMU (WPI), UTIAS, The University of Tokyo,
Kashiwa, Chiba 277-8583, Japan}
\end{center}
\vspace{1mm}

\begin{abstract}
\noindent
In a predictive modular invariant theory of flavour there should exist
% an interplay and
correlations  between the quark and lepton observables. So far these observables have been analyzed separately, making it impossible to investigate their interconnections. We perform for the first time a joint analysis of quark and lepton observables (22 altogether) in a modular flavour model. The model is based on $2O$ flavour symmetry and, within its class, it is characterized by the minimal number of free parameters (14 real constants).
The joint analysis shows that the model is in good agreement with the experimental data for normal neutrino mass
ordering, while predicting the leptonic Dirac CP-violating (CPV) phase ($\delta_{CP}$), the two Majorana CPV phases ($\eta_1$, $\eta_2$), the lightest neutrino mass ($m_1$) and the effective neutrino masses probed by beta and neutrinoless double beta decay ($m_\beta$ and $m_{\beta\beta}$).
 A detailed comparison of the separate (lepton-only and quark-only) and combined (lepton and quark) fit results shows
 differences in best-fit values and jointly allowed regions, that reflect a nontrivial interplay between
quark and lepton observables in the model. Most importantly, our analysis highlights the existence of significant correlations between various pairs of such observables. For instance, the ratio of the strange and bottom quark masses, $r_{sb}$, is strongly negatively correlated with each of the three lepton mixing angles and with $\delta_{CP}$, $m_1$,  $m_\beta$ and $m_{\beta\beta}$, while being positively correlated with $\eta_1$ and $\eta_2$. These findings, that are missed in separate analyses of quark and lepton flavour sectors, fall within ranges that can be tested by current and future experiments.

\end{abstract}

\end{titlepage}
\setcounter{footnote}{0}
%%%%%%%%%%%%%%%%%%%%%%%
%

{\hypersetup{linkcolor=black}
\tableofcontents
}

\vskip 1cm

\vfill
\clearpage

%%%%%%%%%%%%%%%%%%%%%%%
\section{Introduction}
\label{sec:intro}
%%%%%%%%%%%%%%%%%%%%%%%

\hskip 0.5cm
Unraveling the fermion flavor problem in both the quark and lepton sectors, i.e., understanding the origins of the patterns of quark and lepton masses, mixings, and CP-symmetry violations, is one of the most challenging and unsolved issues in particle physics. Although this problem is quite old and ``solutions'' to some of its specific aspects (such as the fermion mass hierarchies, the smallness of neutrino masses, the pattern of neutrino mixing, etc.) have been proposed in the literature (see, e.g.,~\cite{Feruglio:2015jfa}), a universal, elegant, natural and viable theory of flavour is still lacking. Constructing such a theory would be a major achievement and a breakthrough in particle physics, equivalent in importance to the construction of the Standard Model (SM).

A promising approach to the flavour problem based on modular invariance
% % inspired by the string theory,
% %, which is assumed to
% playing the role of a flavour symmetry
was put forward in 2017 in~\cite{Feruglio:2017spp}
and since 2018 it has been actively explored and developed.
%
% In the last seven years modular symmetry inspired by string theory
% has been used to address the flavour problem of the SM
%\cite{Feruglio:2017spp}.
%
% A step forward towards
% % the solution of the flavour problem
% in this direction
% was made in 2017 in Ref.~\cite{Feruglio:2017spp}, where
% the idea of using modular invariance as a flavour
% symmetry was put forward.
% The idea was further explored in the first half of 2018 in
% \cite{Kobayashi:2018vbk,Penedo:2018nmg,Criado:2018thu},
% where phenomenologically viable lepton flavour models
% based on modular symmetry where successfully constructed
% and, since then, the modular invariance approach
% to the flavour problem has been and continues to be
% intensively investigated and developed with encouraging results.
%
In this approach
% In the modular-invariance approach to the flavour problem
the elements of the Yukawa coupling and fermion mass matrices in the Lagrangian are expressed in terms of modular forms characterized by a level $N$ (a natural number), and by a limited number of coupling constants.
The modular forms are holomorphic functions of a single complex scalar
field $\tau$ (the modulus). Both the modulus $\tau$ and the modular forms
have specific transformation properties under the action of the modular group $\Gamma \equiv SL(2,\mathbb{Z})$. The matter fields are assumed to transform in representations of an inhomogeneous (homogeneous) finite modular group $\Gamma^{(\prime)}_N$, while the modular forms furnish irreducible representations of the same group. For $N\leq 5$, the inhomogeneous finite modular groups $\Gamma_N$ are isomorphic to the permutation groups
$S_3$, $A_4$, $S_4$ and $A_5$ (see, e.g.,~\cite{deAdelhartToorop:2011re}),
and the homogeneous groups $\Gamma^\prime_N$ are isomorphic to
their double covers. These groups are quite extensively used in
flavour model building (see, e.g.,~\cite{Altarelli:2010gt,Ishimori:2012zz,Petcov:2017ggy,Feruglio:2019ybq,Kobayashi:2022moq,Ding:2023htn,Ding:2024ozt}).
The modular symmetry described by the finite modular group $\Gamma^{(\prime)}_N$ plays the role of a flavour symmetry and the theory is assumed to be invariant under the whole modular group $\Gamma$. These symmetries are broken by the vacuum expectation value (VEV) of the modulus $(\tau_\mathit{VEV})$, and it can be the only source of flavour symmetry breaking such that flavons are not needed.
%
% A very appealing feature of this approach is that
% the vacuum expectation value (VEV) of the modulus $\tau$
% can be the only source of flavour
% symmetry breaking, such that flavons are not needed.
%
Another appealing feature of the discussed framework is that $\tau_\mathit{VEV}$ can also be the only source of breaking of CP symmetry~\cite{Novichkov:2019sqv}. After the flavour symmetry is
broken, the modular forms and thus the elements of the Yukawa coupling
and fermion mass matrices get fixed. Correspondingly, the fermion mass matrices exhibit a certain symmetry-constrained flavour structure.

There is no VEV of $\tau$ which preserves the full modular symmetry.
However, as pointed out in~\cite{Feruglio:2017spp,Novichkov:2018ovf}
and exploited for model building in~\cite{Novichkov:2018ovf,Novichkov:2018yse,Novichkov:2018nkm,Okada:2020brs},
for explaining the fermion mass hierachies without fine-tuning in~\cite{Novichkov:2021evw} and in~\cite{Feruglio:2021dte,Petcov:2022fjf,Abe:2023qmr,Petcov:2023vws,deMedeirosVarzielas:2023crv},
and from the point of view of criticality in~\cite{Feruglio:2022koo,Feruglio:2023mii,Ding:2024xhz},
there exist three values of $\tau_\mathit{VEV}$ in the modular group fundamental domain, which do not break the modular symmetry completely. These so-called ``fixed points'' are $\tau_\text{sym}= i,\, \omega,\,i\infty$,
with $\omega\equiv\exp(2\pi i/3)=-\,1/2+ i\,\sqrt{3}/2$ (the ``left cusp''), and, for theories based on $\Gamma_N$ invariance, they preserve $\mathbb{Z}^{S}_2$, $\mathbb{Z}^{ST}_3$, and $\mathbb{Z}^{T}_N$ residual symmetries, respectively\footnote{In the case of the double cover groups $\Gamma^\prime_N$, these residual symmetries are augmented by
$\mathbb{Z}_2^R$~\cite{Novichkov:2020eep}.}.

The approach to the flavour problem based on modular invariance, which appears naturally in string theory, has been formulated in the framework
of supersymmetric (SUSY) theories\footnote{Recently a non-SUSY formulation
has been proposed in~\cite{Qu:2024rns}.}. Within the framework of, e.g., rigid $\mathcal{N}=1$ SUSY, modular invariance is assumed to be a property of
the superpotential, whose holomorphicity combined with the modular invariance
implies that the elements of the Yukawa coupling matrices are holomorphic modular forms and restricts the number of allowed terms.

Following the bottom-up approach, phenomenologically viable and ``minimal''
lepton flavour models based on modular symmetry	(not including flavons)
have been constructed first using the groups $\Gamma_3 \simeq A_4$ and $\Gamma_4 \simeq S_4$ in~\cite{Feruglio:2017spp,Penedo:2018nmg}. Models with flavons and small number of parameters based on $\Gamma_2 \simeq S_3$ and $\Gamma_3 \simeq A_4$ have been proposed in~\cite{Kobayashi:2018vbk,Criado:2018thu}. After these initial studies, the interest in the approach grew significantly. Various aspects of the formalism
have been extensively studied and developed and a large variety of models has been constructed.
% and extensively studied.
This includes\footnote{
% A rather complete list of the articles on modular-invariant
% models of lepton and/or quark flavour, which appeared by March of 2021,
% can be found in~\cite{Novichkov:2021evw}.
The number of publications on the modular-invariance approach
to the lepton and quark flavour problems currently exceeds 250.
We cite here only a representative sample.}:
\begin{enumerate}[label=\roman*),  itemsep=-8pt, topsep=-8pt, parsep=0pt, partopsep=0pt]
\item
lepton flavour models based on the groups
\(\Gamma_4 \simeq S_4\)~\cite{Novichkov:2018ovf,Kobayashi:2019mna,Kobayashi:2019xvz,Ding:2019gof,Wang:2019ovr},
\(\Gamma_3 \simeq A_4\)~\cite{Kobayashi:2018scp,Novichkov:2018yse,Ding:2019zxk,Ding:2019gof,Kobayashi:2019gtp},
\(\Gamma_5 \simeq A_5\)~\cite{Novichkov:2018nkm,Ding:2019xna,Novichkov:2021evw},
\(\Gamma_2 \simeq S_3\)~\cite{Marciano:2024nwm,Okada:2019xqk}
and \(\Gamma_7 \simeq PSL(2,\mathbb{Z}_7)\)~\cite{Ding:2020msi};\\
\item
models of quark flavour~\cite{Okada:2018yrn} and of quark--lepton
unification~\cite{Kobayashi:2018wkl,Okada:2019uoy,Kobayashi:2019rzp,Lu:2019vgm,Okada:2020rjb,Chen:2021zty};\\
\item
models with multiple moduli considered first phenomenologically
in~\cite{Novichkov:2018ovf,Novichkov:2018yse} and
% further studied, e.g.,
% in~\cite{deMedeirosVarzielas:2019cyj,King:2019vhv,Ding:2020zxw};\\
shown to appear naturally in theories with simplectic modular symmetries in~\cite{Ding:2020zxw,Ding:2021iqp} or the direct product of several modular groups~\cite{deMedeirosVarzielas:2019cyj};\\
\item
models in which the formalism of the interplay of modular and
generalised CP (gCP) symmetries, developed and applied first to the
lepton flavour problem in~\cite{Novichkov:2019sqv}, is explored~\cite{Kobayashi:2019uyt,Okada:2020brs,Yao:2020qyy,Wang:2021mkw,Ding:2021iqp,Qu:2021jdy}.\\
\end{enumerate}
Also the formalism of the double cover finite modular groups~\(\Gamma'_N\), to which top-down constructions typically lead (see, e.g.,~\cite{Nilles:2020nnc,Kikuchi:2020nxn} and references therein),
has been developed and viable flavour models have been constructed
for the cases of \(\Gamma'_3 \simeq T'\)~\cite{Liu:2019khw}, \(\Gamma'_4 \simeq S'_4\)~\cite{Novichkov:2020eep,Liu:2020akv} and \(\Gamma'_5 \simeq A'_5\)~\cite{Wang:2020lxk,Yao:2020zml}. In Ref.~\cite{Liu:2021gwa} the framework has been further generalised to arbitrary finite modular groups (i.e.,~those not described by series \(\Gamma_N^{(\prime)}\)).

The problem of modulus stabilisation has also been addressed in~\cite{Kobayashi:2019xvz,Abe:2020vmv,Ishiguro:2020tmo,Novichkov:2022wvg,Knapp-Perez:2023nty}.
More recently it was shown that the modular invariance approach to flavour
allows for axion-less solutions of the strong CP problem~\cite{Feruglio:2023uof,Petcov:2024vph,Penedo:2024gtb,Feruglio:2024ytl},
and that the modulus can play the role of an inflaton in the slow-roll
inflation scenarious~\cite{Ding:2024neh,King:2024ssx}.

It is hoped that the results obtained in the bottom-up modular-invariant approach to the lepton and quark flavour problems will eventually connect with top-down results based on UV-complete theories (see, e.g., ~\cite{Kobayashi:2018rad,Kobayashi:2018bff,Baur:2019kwi,Baur:2019iai,Kobayashi:2020hoc,Ohki:2020bpo,Nilles:2020kgo,Nilles:2020tdp,Baur:2020jwc,Ishiguro:2020nuf,Hoshiya:2020hki,Baur:2020yjl,Baur:2021bly,Nilles:2021glx,Kikuchi:2021ogn}.
% \cite{Kobayashi:2018rad, Kobayashi:2020hoc,Abe:2020vmv,Ohki:2020bpo, Nilles:2020kgo, Nilles:2020tdp,Ishiguro:2020nuf,Ishiguro:2020tmo, Baur:2020yjl, Almumin:2021fbk,Baur:2021bly}),
% based on UV-complete theories.
Recently in~\cite{Baur:2022hma} the first string-derived flavor
model with realistic phenomenology, based on ''eclectic''  symmetry,
one component of which is the $\Gamma^\prime_3 = A^\prime_4 \equiv T^\prime$ modular symmetry and another -- the $\Delta (54)$ flavour symmetry -- was constructed. Lepton flavour models employing the minimal ``eclectic'' symmetries
% \ensuremath{\Delta}(27) \ensuremath{\rtimes} S$_{3}$
$\Delta (27)\rtimes T'$~\cite{Ding:2023ynd}, $\Delta (27)\rtimes S_3$~\cite{Li:2023dvm} and $Q_{8}\rtimes S_3$~\cite{Li:2024pff} have been proposed.

So far,  phenomenologically viable modular-invariant models for the lepton or quark flavor sectors have been constructed with ``minimal'' sets of free parameters (two of them being the real and imaginary parts of the complex $\tau_\mathit{VEV}$) such as:
\begin{enumerate}[label=\roman*),  itemsep=0pt, topsep=0pt, parsep=0pt, partopsep=0pt]
\item for the lepton sector with massive Majorana neutrinos (12 observables), in terms of 6 real parameters (4 real couplings and  $\tau_\mathit{VEV}$) in~\cite{Ding:2022nzn},
	and of 7 parameters in~\cite{Novichkov:2018ovf};
\item for the quark sector (10 observables)	in terms of  9 real parameters (7 real couplings and  $\tau_\mathit{VEV}$)
	\cite{Liu:2020akv,Qu:2021jdy},
\item for both lepton and quark flavour sectors (22 observables) by using
14 real parameters (12 real couplings and $\tau_\mathit{VEV}$)~\cite{Ding:2023ydy}.
\end{enumerate}

Within its class, the latter model --- based on a modular $2O$	flavour symmetry \cite{Ding:2023ydy} --- is unique, in that it uses the smallest set of free parameters at present (14 real constants) to describe jointly the quark and lepton observables (22 altogether).

Given the fact that in ``minimal'' flavour models the number of parameters is smaller than the number of the
described observables and that all observables are linked by $\tau_\mathit{VEV}$, there are unusual correlations between observables that are present uniquely in flavour theories  based on modular invariance.	Which observables are correlated and the type of correlations they exhibit depend on the model.
	
More specifically, in ``minimal'' models for the lepton flavor sector, the predicted values of the CPV phases
$\delta_{CP}$, $\eta_1$ and $\eta_2$ and of $m_{\beta\beta}$ may be correlated with $\sin^2\theta_{23}$, while the prediction for the sum of neutrino masses $\sum_i m_i$ may be correlated with the predicted values of $\delta_{CP}$ etc.~\cite{Novichkov:2018ovf}.\footnote{Note, for example, that $m_{\beta\beta}$ does not depend explicitly on $\sin^2\theta_{23}$ and that the CPV phases, $\sum_i m_i$ and $\sin^2\theta_{23}$ are physically very different observables.}

So far, in modular-invariance models for the quark and lepton flavor sectors,	the corresponding observables have been	analyzed separately, preventing the study of their expected interconnections.
We perform for the first time a joint analysis of both the quark and lepton observables in a specific modular flavour model, which allows us to investigate their interplay and correlations. In particular, the analysis is performed within the previously mentioned model with modular $2O$ flavour symmetry constructed in~\cite{Ding:2023ydy}, in terms of 14 free parameters and 22 observables.
The main aim of our study is not the statistical analysis of the $2O$ model itself, but the joint analysis of both the quark and lepton observables, in order to highlight their interconnections in a representative
minimal model based on modular invariance.

The paper is organized as follows: in Section~\ref{sec:model}, we introduce
the theoretical framework of modular symmetries and the finite group $2O$, followed by the detailed construction of the flavor model.
In Section~\ref{sec:analysis}, we perform the combined analysis of quark and lepton observables, presenting the results of our fit to experimental data, the predictions of the model for the unknown lepton parameters
and, most importantly, the interplay and the correlations between the quark and lepton observables emerging from the model.
Finally, in Section~\ref{sec:conclusions}, we discuss the implications of our
findings and provide concluding remarks on the
% predictions of the model including predictive power of the model and its
testability of the obtained predictions in future experiments. Additional technical details and derivations are presented in the Appendix.

\section{Predictive model based on the modular symmetry $2O$}\label{sec:model}

In the following, we shall present the predictive model for quarks and leptons based on the finite modular group $2O$~\cite{Ding:2023ydy}.
The group $2O$, its representations and modular forms relevant for our analysis are described in Appendix~\ref{app:Modular-forms}. The generalized CP (gCP) symmetry is imposed on the Lagrangian of the model and consequently all coupling constants are constrained to be real. The experimentally measured masses and mixing angles of both quarks and leptons can be described by 12 real parameter and a common complex modulus $\tau$.

\subsection{Lepton sector \label{sec:model-lepton-sector}}

In the considered model, the neutrino masses are generated by the type-I seesaw mechanism and the light massive neutrinos are Majorana particles.
% The neutrinos are assumed to be Majorana particles, and the masses
% are generated by the type-I seesaw mechanism.
Both left-handed (LH) lepton doublets and right-handed (RH) neutrino singlets
transform as triplets $\bm{3}$ under $2O$, the RH charged lepton $e^c$ and $\mu^c$ are embedded into a doublet of $2O$, while $\tau^c$ is assumed to be a non-trivial $2O$ singlet $\mathbf{1'}$. The representation assignments and the modular weights of the lepton fields are given by
\begin{eqnarray}\nonumber
&&L\sim \mathbf{3}\,,~~E_{D}^{c}\equiv (e^{c},\mu^{c}) \sim \mathbf{\widehat{2}'}\,,~~\tau^{c}\sim \mathbf{1'}\,,~~N^{c}\sim \mathbf{3}\,,\\
&&k_{L}=-1\,,~~k_{E_{D}^{c}}=6\,,~~k_{\tau^{c}}=5\,,~~k_{N^{c}}=1\,.
\end{eqnarray}
The Higgs fields $H_{u}$ and $H_d$ are invariant singlets of $2O$ with zero modular weights. Thus one can read out the modular invariant superpotentials for charged lepton and neutrino masses as follow,
\begin{eqnarray}\nonumber
\mathcal{W}_{E}&=&g^{E}_{1} \left( E_{D}^{c}L\right)_{\mathbf{\widehat{2}'}}Y^{(5)}_{\mathbf{\widehat{2}'}}H_{d}+g^{E}_{2} \left(E_{D}^{c}L\right)_{\mathbf{\widehat{4}}}Y^{(5)}_{\mathbf{\widehat{4}}}H_{d} +g^{E}_{3}\left(\tau^{c}L\right)_{\mathbf{3'}}Y^{(4)}_{\mathbf{3'}}H_{d}\,,\\
\mathcal{W}_{\nu}&=&gH_{u}(N^{c}L)_{\mathbf{1}}+ \Lambda  \left(N^{c}N^{c}\right)_{\mathbf{2}}Y^{(2)}_{\mathbf{2}}\,.
\end{eqnarray}
The charged lepton mass matrix, the neutrino Dirac mass matrix and the heavy Majorana neutrino mass matrix read:
\begin{eqnarray} \nonumber
M_{E}&=&\left( \begin{array}{ccc}
-g^{E}_{1}Y^{(5)}_{\mathbf{\widehat{2}'},2}-\sqrt{2}g^{E}_{2}Y^{(5)}_{\mathbf{\widehat{4}},3} ~&  \sqrt{3}g^{E}_{2}Y^{(5)}_{\mathbf{\widehat{4}},1} & ~\sqrt{2}g^{E}_{1}Y^{(5)}_{\mathbf{\widehat{2}'},1}+g^{E}_{2}Y^{(5)}_{\mathbf{\widehat{4}},4}\\
-g^{E}_{1}Y^{(5)}_{\mathbf{\widehat{2}'},1}+\sqrt{2}g^{E}_{2}Y^{(5)}_{\mathbf{\widehat{4}},4} ~&-\sqrt{2}g^{E}_{1}Y^{(5)}_{\mathbf{\widehat{2}'},2}+g^{E}_{2}Y^{(5)}_{\mathbf{\widehat{4}},3} & ~-\sqrt{3}g^{E}_{2}Y^{(5)}_{\mathbf{\widehat{4}},2}  \\
g^{E}_{3} Y^{(4)}_{\mathbf{3'},1}~& g^{E}_{3} Y^{(4)}_{\mathbf{3'},3} & ~g^{E}_{3} Y^{(4)}_{\mathbf{3'},2}  \\
\end{array} \right) v_{d}\,,\\
M_D &=& g\begin{pmatrix}
1 & ~0~ & 0 \\
0 & ~0~ & 1 \\
0 & ~1~ &0 \\
\end{pmatrix}v_{u} \,,\quad
M_N =\begin{pmatrix}
-2Y^{(2)}_{\mathbf{2},1} ~&~ 0 ~&~0 \\
 0 ~&~ \sqrt{3}Y^{(2)}_{\mathbf{2},2}  ~&~ Y^{(2)}_{\mathbf{2},1}  \\
 0 ~&~ Y^{(2)}_{\mathbf{2},1} ~&~\sqrt{3}Y^{(2)}_{\mathbf{2},2} \\
\end{pmatrix} \Lambda\,,
\end{eqnarray}
where $Y^{(k)}_{\bm{r}, i}$ denotes the $i$-th component of the modular form multiplet $Y^{(k)}_{\bm{r}}(\tau)$. The light neutrino mass matrix is given by the seesaw formula,
\begin{equation}
M_{\nu}=-M^T_DM^{-1}_NM_D=\frac{g^2v^2_{u}}{\Lambda}\left( \begin{array}{ccc} \frac{1}{2Y^{(2)}_{\mathbf{2},1}} ~& 0 ~& 0 \\
0 ~& \frac{\sqrt{3}Y^{(2)}_{\mathbf{2},2}}{Y^{(2)2}_{\mathbf{2},1}-3Y^{(2)2}_{\mathbf{2},2}} ~& -\frac{Y^{(2)}_{\mathbf{2},1}}{Y^{(2)2}_{\mathbf{2},1}-3Y^{(2)2}_{\mathbf{2},2}} \\
0 ~& -\frac{Y^{(2)}_{\mathbf{2},1}}{Y^{(2)2}_{\mathbf{2},1}-3Y^{(2)2}_{\mathbf{2},2}} ~& \frac{\sqrt{3}Y^{(2)}_{\mathbf{2},2}}{Y^{(2)2}_{\mathbf{2},1}-3Y^{(2)2}_{\mathbf{2},2}} \end{array} \right)\,,\label{eq:equation}
\end{equation}
which gives rise to the three light neutrino masses,
\begin{equation}
m_{1}=\frac{1}{|2Y^{(2)}_{\mathbf{2},1}|}\frac{g^2v^2_{u}}{\Lambda}\,,~~
m_{2}=\frac{1}{|Y^{(2)}_{\mathbf{2},1}-\sqrt{3}Y^{(2)}_{\mathbf{2},2}|}\frac{g^2v^2_{u}}{\Lambda}\,,~~
m_{3}=\frac{1}{|Y^{(2)}_{\mathbf{2},1}+\sqrt{3}Y^{(2)}_{\mathbf{2},2}|}\frac{g^2v^2_{u}}{\Lambda}\,.\label{eq:equation2}
\end{equation}
The gCP symmetry enforces all couplings to be real. The charged lepton mass matrix $M_{E}$ depends three parameters $g^{E}_{1}$, $g^{E}_{2}$, $g^{E}_{3}$ which can be adjusted to accommodate the hierarchical masses of electron, muon and tau. The light neutrino matrix $M_{\nu}$ is completely determined by the modulus $\tau$ and the overall scale $g^2v^2_u/\Lambda$. Hence this model only involves $6$ free real input parameters including $\text{Re}(\tau)$ and $\text{Im}(\tau)$. It is the modular invariant model with the smallest number of free parameters needed to describe the quark and lepton flavor sectors, within the current literature. This makes it particularly interesting for the purposes of this work: the fewer the free parameters, the stronger the expected correlations.

\subsection{\label{sec:model-quark-sector}Quark sector}

The third generation of quark fields are singlets of the $2O$ modular group in order to account for their heavy masses, while the first two generations of quark fields are doublets of $2O$. The assignments of quarks are given by
\begin{eqnarray}\nonumber
&&Q_{D}\equiv (Q_{1},Q_{2})^{T}\sim \mathbf{2}\,,~~Q_{3}\sim \mathbf{1'}\,,~~U_{D}^{c}\equiv (u^{c},c^{c})\sim \mathbf{\widehat{2}'}\,,\\
\nonumber
&&t^{c}\sim \mathbf{1'}\,,~~D_{D}^{c}\equiv (d^{c},s^{c}) \sim \mathbf{2}\,,~~b^{c}\sim \mathbf{1'}\,,\\
&& k_{Q_{D}}=3-k_{U_{D}^{c}}=k_{Q_{3}}=6-k_{t^{c}}=6-k_{D_{D}^{c}}=-k_{b^{c}}\,.
\end{eqnarray}
Hence the modular invariant quark superpotentials read:
\begin{eqnarray}\nonumber
\mathcal{W}_{u}&=&g^{u}_{1}\left( U_{D}^{c}Q_{D} \right)_{\widehat{\mathbf{4}}}Y_{\widehat{\mathbf{4}}}^{(3)}H_{u} + g^{u}_{2}\left( t^{c}Q_{D}\right)_{\mathbf{2}}Y^{(6)}_{\mathbf{2}}H_{u}+ g^{u}_{3}\left( t^{c}Q_{3}\right)_{\mathbf{1}}Y^{(6)}_{\mathbf{1}}H_{u}\,,\\ \nonumber
\mathcal{W}_{d}&=&g^{d}_{1}\left( D_{D}^{c}Q_{D} \right)_{\mathbf{1}}Y^{(6)}_{\mathbf{1}}H_{d}+g^{d}_{2}\left( D_{D}^{c}Q_{D} \right)_{\mathbf{1'}}Y^{(6)}_{\mathbf{1'}}H_{d}+g^{d}_{3}\left( D_{D}^{c}Q_{D} \right)_{\mathbf{2}}Y^{(6)}_{\mathbf{2}}H_{d}\\
&~&+g^{d}_{4}\left( D_{D}^{c}Q_{3} \right)_{\mathbf{2}}Y^{(6)}_{\mathbf{2}}H_{d}+g^{d}_{5}\left( b^{c}Q_{3} \right)_{\mathbf{1}}H_{d} \,.
\end{eqnarray}
After symmetry breaking, the quark mass matrices take following forms
\begin{eqnarray}\nonumber
M_{u}&=&\left(\begin{array}{ccc}
g^{u}_{1}Y_{\widehat{\mathbf{4}},3}^{(3)} & -g^{u}_{1}Y_{\widehat{\mathbf{4}},2}^{(3)} & 0 \\
g^{u}_{1}Y_{\widehat{\mathbf{4}},4}^{(3)} & g^{u}_{1}Y_{\widehat{\mathbf{4}},1}^{(3)} & 0 \\
-g^{u}_{2} Y_{\mathbf{2,2}}^{(6)} & g^{u}_{2} Y_{\mathbf{2,1}}^{(6)} & g^{u}_{3} Y_{\mathbf{1}}^{(6)} \\
\end{array}
\right)v_{u} \,,\\
M_{d}&=&\left(
\begin{array}{ccc}
g^{d}_{1}Y_{\mathbf{1}}^{(6)}-g^{d}_{3} Y_{\mathbf{2,1}}^{(6)} & g^{d}_{2} Y_{\mathbf{1'}}^{(6)}+g^{d}_{3} Y_{\mathbf{2,2}}^{(6)} & -g^{d}_{4} Y_{\mathbf{2,2}}^{(6)} \\
g^{d}_{3} Y_{\mathbf{2,2}}^{(6)}-g^{d}_{2} Y_{\mathbf{1'}}^{(6)} & g^{d}_{3} Y_{\mathbf{2,1}}^{(6)}+g^{d}_{1}Y_{\mathbf{1}}^{(6)} & g^{d}_{4} Y_{\mathbf{2,1}}^{(6)} \\
0 & 0 & g^{d}_{5} \\
\end{array}
\right)v_{d} \,.
\end{eqnarray}
We see that the model has 8 real parameters: $g^{u}_{1}$, $g^{u}_{2}$, $g^{u}_{3}$ in the up-type quark sector and $g^{d}_{1}$, $g^{d}_{2}$, $g^{d}_{3}$, $g^{d}_{4}$, $g^{d}_{5}$ in the down-type quark sector.  The complex modulus $\tau$ in the quark and lepton sectors is the same one, so the total number of real free parameters is $6+10-2=14$. Thus this model is
very predictive, since it uses 14 real free parameters to describe the 22 masses and mixing parameters of quarks and leptons.

\section{Correlated flavor structure: quark and lepton analysis}
\label{sec:analysis}

In this section, we present the results of the numerical analysis of the
$2O$ model with the aim of assessing the interplay and correlations between the quark and lepton observables. For this purpose we perform first separate analyses of the lepton sector and of the quark sector, which is followed by a joint analysis of both the quark and lepton sectors.

We recall that the theoretical model is characterized by 14 real parameters,
collectively denoted as $\mathbf{P}$:
%%%%%%%%%%%%%%%%%%%%%%%%%%%%%
\renewcommand{\arraystretch}{1.5} % Increase row spacing locally
\begin{equation}
\text{P}_i = \left\{
\begin{array}{lll}
\text{Re}(\tau),\text{Im}(\tau)& i= 1,2& \text{modulus} , \\
\displaystyle  g^E_1\, v_d, \frac{g^E_2}{g^E_1}, \frac{g^E_3}{g^E_1},\frac{g\, v_u}{\sqrt{\Lambda}}& i= 3,4,5,
6& \text{leptons} ,\\
\displaystyle g^u_1\, v_u, \frac{g^u_2}{g^u_1}, \frac{g^u_3}{g^u_1},
g^d_1\, v_d, \frac{g^d_2}{g^d_1}, \frac{g^d_3}{g^d_1}, \frac{g^d_4}{g^d_1}, \frac{g^d_5}{g^d_1}& i= 7,.\cdots14& \text{quarks} .
\end{array}
\right.\label{eq:3.1}
\end{equation}
\renewcommand{\arraystretch}{1} % Increase row spacing locally
Out of the 14 parameters in $\mathbf{P}$, only four are dimensional, and we find convenient to use the following units
\renewcommand{\arraystretch}{1.5} % Increase row spacing locally
\begin{equation}
\displaystyle g^E_1 v_d\  (\text{MeV}),\quad \frac{g\, v_u}{\sqrt{\Lambda}}\ (\sqrt{\text{eV}})\, \quad g^u_1 v_u \  (\text{GeV}),\quad
g^d_1 v_d\  (\text{GeV})\ .\label{eq:3.2}
\end{equation}
\renewcommand{\arraystretch}{1} % Increase row spacing locally
The choices in equations~\ref{eq:3.2} are motivated by the expressions for the Yukawa superpotentials,
$\mathcal{W}_{u}$ and $\mathcal{W}_{d}$.
After symmetry breaking, the coefficients $g_1^u\, v_u$ and $g_1^d\, v_d$ are
factored out and control the up and down quark masses, respectively.
Similarly, in the neutrino sector, we factor out the terms $g_1^E v_d$ and $g v_u/\sqrt{\Lambda}$, which set the
scales of the charged lepton and neutrino masses, respectively.
Since we are also interested in studying the lepton and quark sectors separately, it is useful to define the following subsets of parameters:
\begin{align}
\mathbf{P}_{\text{leptons}} &= (\tau,\displaystyle  g^E_1\, v_d, \frac{g^E_2}{g^E_1}, \frac{g^E_3}{g^E_1},\frac{g\, v_u}{\sqrt{\Lambda}})\ ,\\
\mathbf{P}_{\text{quarks}} &= (\tau,\displaystyle  \displaystyle g^u_1\, v_u, \frac{g^u_2}{g^u_1}, \frac{g^u_3}{g^u_1},
g^d_1\, v_d, \frac{g^d_2}{g^d_1}, \frac{g^d_3}{g^d_1}, \frac{g^d_4}{g^d_1}, \frac{g^d_5}{g^d_1})\ ,
\end{align}
which consist of six and ten parameters, respectively, and share the complex modulus $\tau$. The possible values of the parameters P$_i$ are obtained from the fit to 18 observables that we collectively group in the vector $\mathbf{O}$:
\renewcommand{\arraystretch}{1.5} % Increase row spacing locally
\begin{equation}
\text{O}_i = \left\{
\begin{array}{lll}
\displaystyle \sin^2\theta_{12}, \sin^2\theta_{23}, \sin^2\theta_{13}& i= 1,2,3& \nu\text{ mixing} , \\
\displaystyle  \delta m^2, \Delta m^2& i= 4,5& \nu\text{ masses}  , \\
\displaystyle r_{e\mu}, r_{\mu\tau},m_\tau& i = 6,7,8& \text{charged lepton masses}, \\
\theta_{12}^q, \theta_{23}^q, \theta_{13}^q, \delta_{CP}^q& i= 9,10,11,12& \text{quark mixing} , \\
\displaystyle r_{uc}, r_{ct} ,r_{ds} ,r_{sb}, m_t, m_b&
i= 13,\dots,18 & \text{ quark masses}.
\end{array}
\right.\label{eq:3_5}
\end{equation}
\renewcommand{\arraystretch}{1} % Increase row spacing locally
The quantities $r_{ab}$ in~\ref{eq:3_5} denote the ratios of the masses of particle $a$ to particle $b$ (either charged leptons or quarks). Consequently, we have 8 observables related to leptons and 10 observables related to quarks. The two neutrino squared mass differences in~\ref{eq:3_5} are $\Delta m^2=m^2_3-{(m^2_1+m^2_2})/2$ and $\delta m^2=m^2_2-m^2_1$.

We choose not to include the leptonic Dirac phase $\delta_{CP}$ in the fit due to its currently large experimental uncertainty~\cite{Capozzi:2021fjo}. Additionally, in the absence of a direct measurement of neutrino masses from beta decay or neutrinoless double-beta decay, the mass of the lightest neutrino, $m_1$, remains unknown. As a result, we impose two fewer constraints in the lepton sector compared to the quark sector.

The experimental values of the leptonic observables and their $1\sigma$ uncertainties are summarized in Table~\ref{tab:tab1}.
The values refer to the phenomenologically viable case of a neutrino mass spectrum with normal ordering (NO), since the IO case is	incompatible with the minimal $2O$ model as reported in~\cite{Ding:2023ydy}.
The neutrino mixing angles and squared mass differences are taken from~\cite{Capozzi:2021fjo}, except for $\sin^2\theta_{23}$. Since the octant of $\sin^2\theta_{23}$ is currently uncertain (both octants being allowed at approximately $2\sigma$),
we adopt $\sin^2\theta_{23}=0.5$ as the average experimental value, with the $1\sigma$ uncertainty set to 1/6 of the $\pm3\sigma$ allowed range.
%===========================================================================
\begin{table}[h!]
\centering
\begin{minipage}{\textwidth}
\caption{Neutrino oscillation parameters: best-fit values and 1$\sigma$ allowed range for Normal Ordering. The latter column shows the ``$1\sigma$ fractional accuracy'' for each parameter, defined as 1/6 of the $\pm3\sigma$ range, divided by the best-fit value and expressed in percent. }
\centering
\begin{tabular}{lccc}
\hline
Parameter  & Best fit & $1\sigma$ range & ``$1\sigma$'' (\%) \\
\hline%---------------------------------------------------------------------
$\delta m^2/10^{-5}~\mathrm{eV}^2 $ & 7.36 & 7.21 -- 7.52  & 2.3 \\
%---------------------------------------------------------------------
$\sin^2 \theta_{12}/10^{-1}$ &  3.03 & 2.90 -- 3.16 & 4.5 \\
%---------------------------------------------------------------------
$\Delta m^2/10^{-3}~\mathrm{eV}^2 $ &  2.485 & 2.454 -- 2.508 & 1.1 \\
%---------------------------------------------------------------------
$\sin^2 \theta_{13}/10^{-2}$ &  2.23 & 2.17 -- 2.30 & 3.0 \\
%---------------------------------------------------------------------
$\sin^2 \theta_{23}/10^{-1}$ & 5.0 & 4.69 -- 5.31   &6.7 \\
%---------------------------------------------------------------------
\hline\label{tab:tab1}
\end{tabular}
\end{minipage}
\end{table}
%===========================================================================
%%%%%%%%%%%%%%%%%%%%%%%%%%

For the case of NO and small $\tan\beta=5$ adopted in the present work,
the renormalization group (RG) effects on the neutrino mass ratios and mixing
angles are known to be negligible to a good approximation~\cite{Antusch:2005gp,Antusch:2013jca}.
For completeness, we performed numerical evaluations of RG corrections
using the state-of-the-art RG software
package \texttt{REAP}~\cite{Antusch:2005gp,Antusch:2013jca} and \texttt{RGE++}
~\cite{Deppisch:2020aoj}, applied to our input model. We find that RG effects
remain at the percent level on mass squared differences and at the
subpercent level on mixing angles, consistent with the theoretical accuracy
estimated in~\cite{Antusch:2013jca}, within the current REAP approximations.
Thus, RG corrections to neutrino mass-mixing parameters can be neglected for our purposes.

However, the charged fermion masses and quark mixing parameters
have been extrapolated to the GUT scale~\cite{Antusch:2013jca}.
In particular we assume:
%%%%%%%%%%%%%%%%%%%%
\begin{align}
&m_{\tau}  = 1.293 \pm 0.007\ \text{GeV}\ ,\nonumber\\
& r_{e\mu} = \frac{m_e}{m_{\mu}} = 0.00474 \pm 0.00004\ ,\nonumber\\
& r_{\mu\tau} = \frac{m_{\mu}}{m_{\tau}} = 0.0588 \pm 0.0005\ .
\end{align}
%%%%%%%%%%%%%%%%%%%%%%%%%%%%%
Table~\ref{tab:tab2} shows the experimental measurements of the masses and mixing parameters in the quark sector.
%%%%%%%%%%%%%%%%%%%%%%%%%%%%%%%
\begin{table}[h!]
\centering
\begin{minipage}{\textwidth}
\caption{\label{tab:tab2} The best fit values $\mu_i$ and $1\sigma$ uncertainties of the quark parameters when evolved to the GUT scale as calculated in~\cite{Antusch:2013jca}, with the SUSY breaking scale $M_{\text{SUSY}}=10$ TeV and $\tan\beta=5$, where the error widths represent $1\sigma$ intervals.}
\centering
\begin{tabular}{lclc}
\hline
Parameters & $\mu_i\pm1\sigma$ & Parameters & $\mu_i\pm1\sigma$ \\ \hline
$m_{t}/\text{GeV}$ & $89.213 \pm 2.219$ &  $m_{s}/m_{b}$ & $0.0182 \pm 0.0010$ \\
$m_{u}/m_{c}$ & $0.00193 \pm 0.00060$ & $\theta_{12}^{q}$ & $0.2274 \pm 0.0007$ \\
$m_{c}/m_{t}$ & $0.00280 \pm 0.00012$ & $\theta_{13}^{q}$ & $0.00349 \pm 0.00013$\\
$m_{b}/\text{GeV}$ & $0.965 \pm 0.011$ & $\theta_{23}^{q}$ & $0.0400 \pm 0.0006 $  \\
$m_{d}/m_{s}$ & $0.0505 \pm 0.0062$ & $\delta_{CP}^{q}/^{\circ}$ & $69.21 \pm 3.11 $\\
\hline
\end{tabular}
\end{minipage}
\end{table}
%%%%%%%%%%%%%%%%%%%%%%%%%%%%%%%%%%%%%%

We define $\mathbf{O}^{\text{exp}}$ and $\boldsymbol{\sigma}^{\text{exp}}$ as the vectors of experimental measurements and 1$\sigma$ uncertainties, respectively, for the lepton and quark observables reported in Tables~\ref{tab:tab1} and~\ref{tab:tab2}, in the same order as introduced in the definitions of equation~\ref{eq:3.1}. To assess whether the minimal $2O$ model can accommodate the experimental measurements, we compute the theoretical predictions for the observables $\mathbf{O}^\text{theo}(\mathbf{P})$ at each point $\mathbf{P}$ in the space of Lagrangian parameters, and then compare these predictions with the experimental values using the $\chi^2$ function:
\begin{equation}
\chi^2_\text{comb}(\mathbf{P}) = \sum_{i=1}^{18} \left(\frac{\text{O}^\text{theo}_i(\mathbf{P}) - \text{O}^\text{exp}_i}{\sigma^\text{exp}_i}\right)^2\ .\label{eq:equation3}
\end{equation}
Before performing a comprehensive analysis of the model using the $\chi^2$ function~\ref{eq:equation3}, we first examine the model separately in the lepton and quark sectors. This is done through two separate $\chi^2$ functions:
%%%%%%%%%%%%%%%%%%%%%%%%%%%%%%%%%
\begin{align}
&\chi^2_\text{leptons}(\mathbf{P}_\text{leptons}) = \sum_{i=1}^{8} \left(
\frac{\text{O}^\text{theo}_i(\mathbf{P}_\text{leptons}) - \text{O}^\text{exp}_i}{\sigma^\text{exp}_i}\right)^2\ ,\\
&\chi^2_\text{quarks}(\mathbf{P}_\text{quarks}) = \sum_{i=9}^{18} \left(\frac{\text{O}^\text{theo}_i(\mathbf{P}_\text{quarks}) - \text{O}^\text{exp}_i}{\sigma^\text{exp}_i}\right)^2\ ,
\end{align}
%%%%%%%%%%%%%%%%%%%%%%%%%%%%%%%%
where the combined $\chi^2$ for both leptons and quarks is simply the sum of the two:
%%%%%%%%%%%%%%%%%%%%%%%%%%%%%%
\begin{equation}
\chi^2_\text{comb} = \chi^2_\text{leptons} + \chi^2_\text{quarks}\ ,
\end{equation}
%%%%%%%%%%%%%%%%%%%%%%%%%
with the parameter $\tau$ shared between both sectors\footnote{We will often use in what follows$\tau$ for the  VEV of $\tau$.}.
. We perform the $\chi^2$ statistical analysis using a Markov Chain Monte Carlo (MCMC) method~\cite{cobaya1,cobaya2,Lewis:2019xzd} to obtain the best-fit points $\mathbf{P}^\text{bf}$, $\mathbf{P}_\text{leptons}^\text{bf}$, and $\mathbf{P}_\text{quarks}^\text{bf}$ in the space of the Lagrangian parameters, as well as the allowed regions around these best fits. We vary $\tau$ in the region where $\text{Re}(\tau) \leq 0$, which corresponds to half of the fundamental domain $\mathcal{F}$, defined as:
%%%%%%%%%%%%%%%%%%%%%%%%%%%%
\begin{equation}
\mathcal{F}\ :\ \ |\text{Re}(\tau)| \leq \frac{1}{2}\ , \ \ \text{Im}(\tau)>0\ ,\ \ |\tau|\geq 1\ .
\end{equation}
%%%%%%%%%%%%%%%%%%%%%%%%%%%%
This restriction is motivated by the fact that allowing the model parameters $\mathbf{P}$ to vary symmetrically around the origin would result in redundancy due to the gCP symmetry, which imposes a reflection symmetry about the imaginary axis of $\tau$. Consequently, we assume that all the Lagrangian parameters are positive, with the exception of $\text{Re}(\tau)$, without loss of generality.

We perform three separate analyses: one for the lepton sector, one for the quark sector, and a combined analysis.
In each of these analyses the number of measured observables equals or exceeds the number of free parameters in the fit. In particular, the quark-only, lepton-only, and quark+lepton analyses involve respectively 10, 6 and 14 free parameters on one hand, and 10, 8 and 18 input data on the other hand. Therefore, while for the quark-only analysis there are as many experimental data as theory parameters, in the lepton-only and quark+lepton analyses the data overconstrain the model. In any case, it is not at all trivial that the model can successfully describe the quark and lepton sectors, either separately or in combination. Conversely, a successful description can not only lead to well-defined predictions for (still) unknown lepton observables, but also to interesting links between all lepton and quark observables, as shown below. In the following figures, we shall often superimpose the separate and combined quark and lepton results, for the sake
of visual comparison and of graphical economy.

\begin{figure}[h!]
\centering
\includegraphics[width=0.7\linewidth]{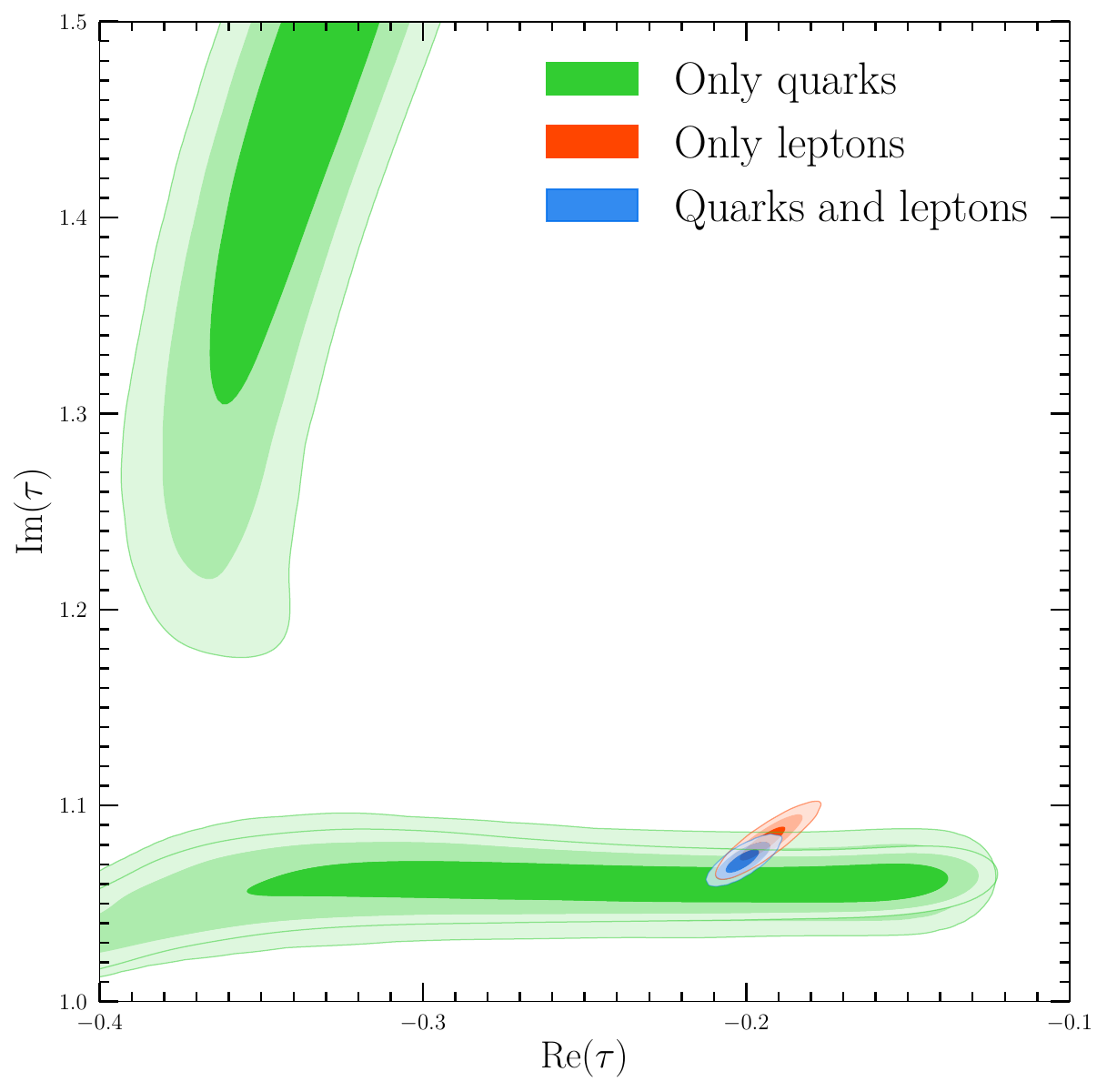}
\caption{Allowed regions in the $\tau$ plane from the lepton-only (red), quark-only (green), and combined (blue) analyses. Different color shadings correspond to the 1$\sigma$, 2$\sigma$, and 3$\sigma$ confidence levels.}
\label{fig:2O_FIG_01}
\end{figure}

Figure~\ref{fig:2O_FIG_01} shows the allowed regions for $\tau$ across the three analyses, while the best-fit values and $\chi^2$ breakdown for each observable are provided in Table~\ref{tab:tab3}.

\renewcommand{\arraystretch}{1.2} % Increase row spacing locally
\begin{table}[ht]
\centering
\caption{Observables best-fit values and corresponding $\chi^2$ breakdown values for the combined analysis, lepton-only, and quark-only analyses.}
\resizebox{\textwidth}{!}{%
\begin{tabular}{c|c|c|c|c|c|c}
\hline
\hline
& \multicolumn{2}{c}{\textbf{Combined}} & \multicolumn{2}{|c|}{\textbf{Leptons only}} &
\multicolumn{2}{c}{\textbf{Quarks only}}  \\ \hline
\text{Observable} & \text{Best fit} & \text{$\chi^2$ breakdown} & \text{Best fit} & \text{$\chi^2$ breakdown}  & \text{Best fit} & \text{$\chi^2$ breakdown}  \\ \hline
$\sin^2\theta_{12}$  & 0.344  & 10.2   & 0.329  & 3.99  & -       & -        \\
$\sin^2\theta_{23}$  & 0.508  & 0.0714  & 0.506  & 0.0438  & -       & -        \\
$\sin^2\theta_{13}$  & 0.0231  & 1.41 & 0.0219  & 0.462  & -       & -        \\
$\delta m^2$ (eV$^2$) & $7.28\times 10^{-5}$ & 0.246  & $7.43 \times 10^{-5}$ & 0.183  & -       & -        \\
$\Delta m^2$ (eV$^2$) & 0.00249 & 0.0636  & 0.00248 & 0.0497  & -       & -        \\
$r_{e\mu}$  & 0.00474 & 2.60$\times10^{-5}$ & 0.00474 & 8.18$\times10^{-6}$ & -       & -        \\
$r_{\mu\tau}$ & 0.0586& 4.66$\times10^{-5}$ & 0.0586 & 4.66$\times10^{-8}$ & -       & -        \\
$m_\tau$ (GeV)   & 1.29   & 5.52$\times10^{-8}$ & 1.29   & 2.94$\times10^{-9}$ & -       & -        \\ \hline
$\min(\chi^2_{\text{leptons}})$ & \multicolumn{2}{c|}{11.94} & \multicolumn{2}{c|}{4.74} & \multicolumn{2}{c}{-} \\ \hline
$\theta_{12}^q$ & 0.227 & 5.68$\times10^{-4}$ & -       & -        & 0.227 & 1.52$\times10^{-7}$ \\
$\theta_{13}^q$ & 0.00351 & 0.0146  & -       & -        & 0.00349  & 5.92$\times10^{-5}$ \\
$\theta_{23}^q$ & 0.0395 & 0.948  & -       & -        & 0.0402  & 2.32$\times10^{-7}$ \\
$\delta_{CP}^q$ & 1.23   & 0.221  & -       & -        & 1.21   & 1.63$\times10^{-8}$ \\
$r_{uc}$   & 0.00212  & 0.100  & -       & -        & 0.00196  & 0.0024  \\
$r_{ct}$   & 0.00282  & 2.61$\times10^{-6}$ & -       & -        & 0.00282  & 2.90$\times10^{-8}$ \\
$r_{ds}$   & 0.0505  & 5.84$\times10^{-5}$ & -       & -        & 0.0505  & 3.08$\times10^{-8}$ \\
$r_{sb}$   & 0.0209  & 7.12   & -       & -        & 0.0183  & 2.62$\times10^{-5}$ \\
$m_b$      & 0.965   & 1.67$\times10^{-7}$ & -       & -        & 0.965   & 1.99$\times10^{-8}$ \\
$m_t$      & 89.2  & 9.42$\times10^{-9}$ & -       & -        & 89.2  & 9.52$\times10^{-11}$ \\ \hline
$\min(\chi^2_{\text{quarks}})$ & \multicolumn{2}{c|}{8.40} & \multicolumn{2}{c|}{-} & \multicolumn{2}{c}{0.00248} \\ \hline
$\min(\chi^2_{\text{comb}})$ & \multicolumn{2}{c|}{20.3} & \multicolumn{2}{c|}{-} & \multicolumn{2}{c}{-} \\
\hline
\hline
\end{tabular}}\label{tab:tab3}
\end{table}
\renewcommand{\arraystretch}{1.} % Increase row spacing locally
%%%
%
%
\subsection{Quark-only analysis}

Concerning the quark-only analysis, as follows from the results
presented in Table~\ref{tab:tab3}, the ten quark observables are perfectly reproduced by the model. The extremely low value of the $\chi^2$ (of $\mathcal{O}(10^{-3})$) may be a result of the fact that the number of free parameters equals the number of experimental constraints. Note that
the two disconnected allowed regions for $\tau$ obtained in the analysis are relatively large compared to the region found in the case of the lepton-only analysis, as Figure~\ref{fig:2O_FIG_01} shows. Also note that in the quark analysis, there is a quasi-horizontal line of nearly degenerate solutions located in the center of the lower green allowed region.

\subsection{Lepton-only analysis}

In the lepton-only analysis, the lepton observables are well fitted only in a much smaller portion of the $\tau$ plane, as is evident in Figure~\ref{fig:2O_FIG_01}. In the green region located in the top left of this figure the lepton observables are not well reproduced, while the quark
observables are excellently fitted in the quark-only analysis. The minimum $\chi^2$ found in the lepton-only analysis is approximately 4.7 (see Table~\ref{tab:tab3}) and is dominated by the pull of $\sin^2\theta_{12}$. Indeed, it is worth noting that all the lepton observables are very well fitted, except for for $\sin^2\theta_{12}$ that
deviates from the experimental value by approximately 2$\sigma$.

% \subsection{Combined Analysis of Lepton and Quark Observables}

\subsection{Combined analysis: interplay and correlations between quark
and lepton observables}

\begin{figure}[h!]
\centering
\includegraphics[width=\linewidth]{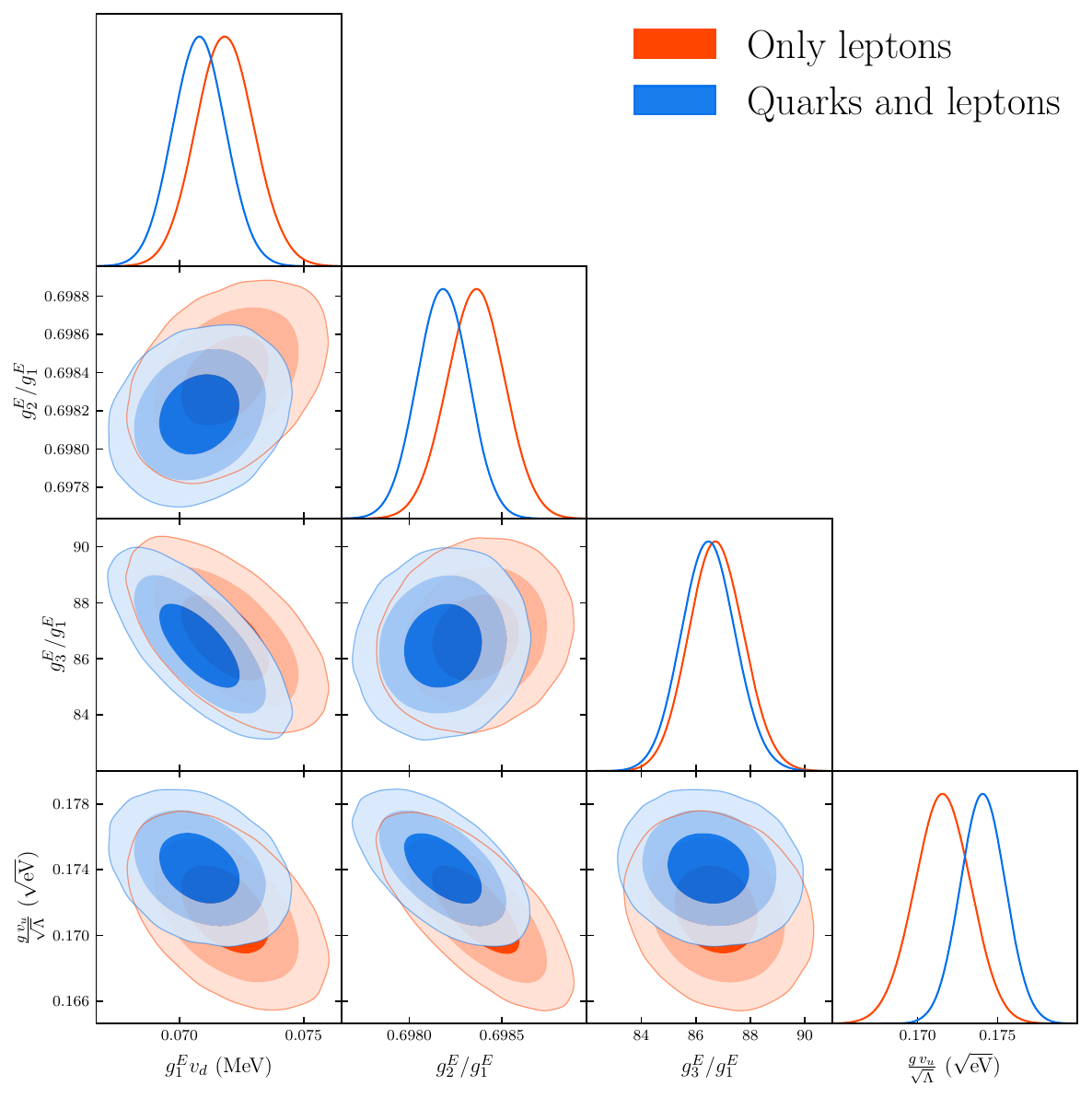}
\caption{Allowed regions for the lepton lagrangian parameters from the lepton-only (red) and combined (blue) analyses. Different color shadings correspond to the 1$\sigma$, 2$\sigma$, and 3$\sigma$ confidence levels.}
\label{fig:2O_FIG_02}
\end{figure}

\begin{figure}[h!]
\centering
\includegraphics[width=\linewidth]{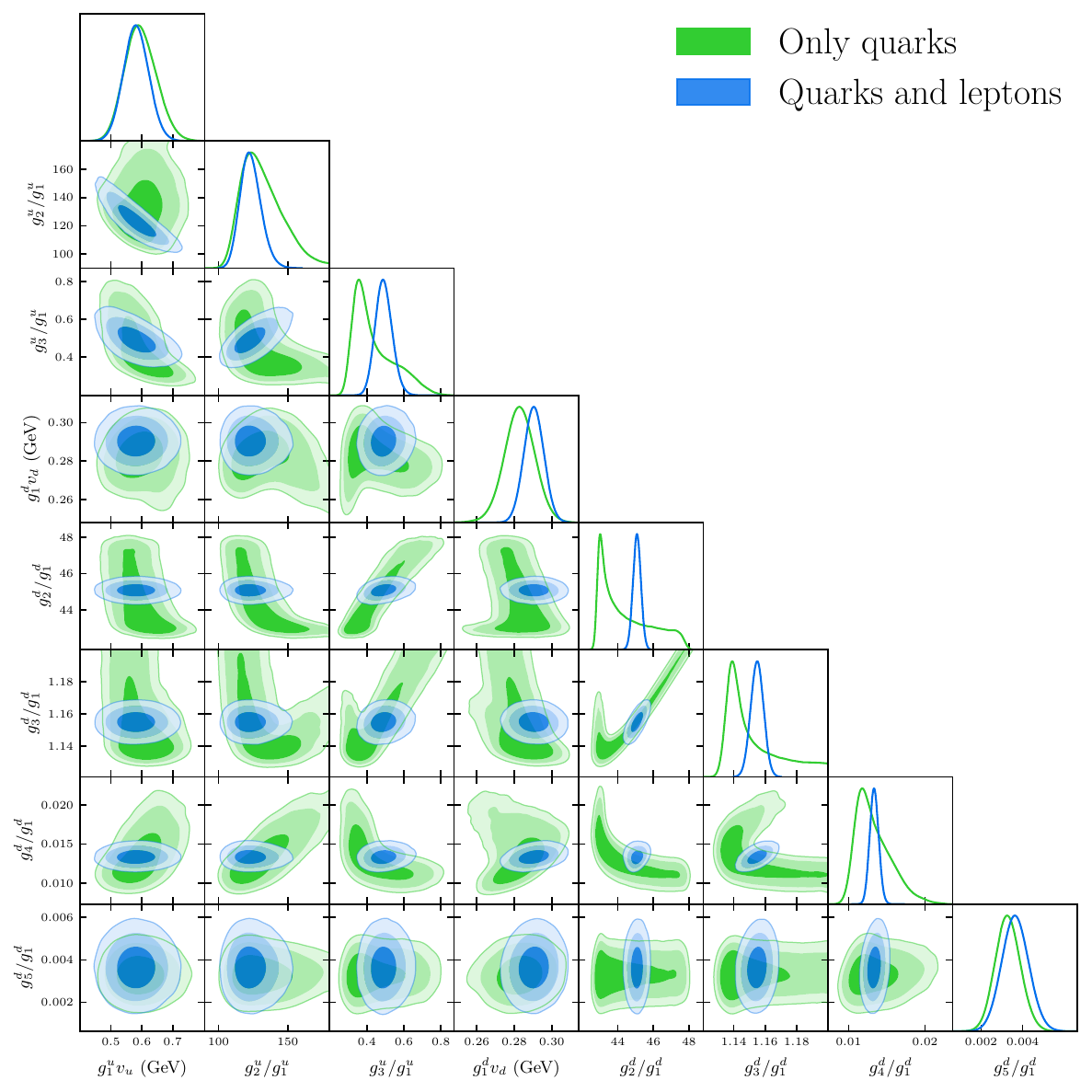}
\caption{Allowed regions for the quark lagrangian parameters from the quark-only (green) and combined (blue) analyses. Different color shadings correspond to the 1$\sigma$, 2$\sigma$, and 3$\sigma$ confidence levels.}
\label{fig:2O_FIG_03}
\end{figure}

The combined analysis of leptons and quarks in the $\tau$ plane reveals a
small allowed region centered around the best-fit point, $\tau=-0.202+1.071i$, with a minimum combined value of $\chi^2_\text{comb}=20.3$.
As seen in Figure~\ref{fig:2O_FIG_01},
this region is located at the intersection of the red lepton-allowed region and the lower green quark-allowed region.
Interestingly, the minimum of the combined $\chi^2_\text{comb}$ is not merely the sum of the lepton and quark minima (approximately 4.74), but significantly higher. This suggests a complex interplay between the lepton and quark sectors, as their respective minima occupy distinct regions of the $\tau$ plane. Consequently, the combined analysis identifies a new global minimum for $\tau$, which minimizes the total $\chi^2$ contribution for all eighteen observables, resulting in the overall $\chi^2_\text{comb}$. Table~\ref{tab:tab4} presents the best-fit values of the Lagrangian parameters for the three different analyses\footnote{At the $\chi^2$ best-fit point, the contribution to the $\chi^2$ from $\sin^2 \theta_{12}$ is computed with a 1$\sigma$ uncertainty  $\Delta \sin^2 \theta_{12}=0.0134\cong 0.013$. From Table I of~\cite{Capozzi:2021fjo}, by estimating the 1$\sigma$ uncertainty as one-sixth of the 3$\sigma$ allowed range, we would obtain an uncertainty of $0.0137\cong0.014$ for $\sin^2\theta_{12}$ and contribution to $\chi^2$ from $\sin^2\theta_{12}$ of 8.6. The value adopted here, namely 10.2, is obtained with the use of $\Delta \sin^2 \theta_{12}=0.013$ and a more accurate interpolation of the $\chi^2$ marginalized profile for $\sin^2 \theta_{12}$, as shown in Figure~3 of~\cite{Capozzi:2021fjo}.}. The corresponding allowed regions are shown in Figures~\ref{fig:2O_FIG_02} and~\ref{fig:2O_FIG_03}.
%%%%%%%%%%%%%%%%%%%%%%%%%%%%%%%%
\renewcommand{\arraystretch}{1.2} % Increase row spacing locally
\begin{table}[ht]
\centering
\caption{Best-fit values of the model parameters for the combined, lepton-only, and quark-only analyses. Note that, in the quark-only analysis, there is a quasi-horizontal line of nearly degenerate minima in the $\tau$ plane, from which a representative point was selected.}
\resizebox{0.8\textwidth}{!}{%
\begin{tabular}{c|c|c|c}
\hline
\hline
\text{Parameter} & \textbf{Combined} & \textbf{Only leptons} & \textbf{Only quarks}  \\ \hline
Re$(\tau)$ & -0.202 & -0.192 & -0.201\\
Im$(\tau)$ & 1.071 &  1.083 & 1.057 \\ \hline
$\displaystyle g^E_1\, v_d$ (MeV) & 0.0707 &0.0714& - \\
$\displaystyle g^E_2/g^E_1$ & 0.698 & 0.716 & -\\
$\displaystyle g^E_3/g^E_1$ & 86.4 & 87.7& -\\
$\displaystyle \frac{g\, v_u}{\sqrt{\Lambda}}$ ($\sqrt{\text{eV}}$) & 0.174 & 0.171& -\\ \hline
$\displaystyle g^u_1\, v_u$ (GeV) & 0.586 & - & 0.578\\
$\displaystyle g^u_2/g^u_1$ & 121.9 & - & 121.1 \\
$\displaystyle g^u_3/g^u_1$ & 0.487 & - &  0.441 \\
$\displaystyle g^d_1\, v_d$ (GeV) & 0.290 & - & 0.330  \\
$\displaystyle g^d_2/g^d_1$ & 45.1 & - &  38.8 \\
$\displaystyle g^d_3/g^d_1$ & 1.16 & - &  0.870 \\
$\displaystyle g^d_4/g^d_1$ & 0.0132 & - & 0.0101  \\
$\displaystyle g^d_5/g^d_1$ & 0.00361 & - & 0.00277  \\
\hline
\hline
\end{tabular} }\label{tab:tab4}
\end{table}
\renewcommand{\arraystretch}{1.} % Increase row spacing locally
%%%%%%%%%%%%%%%%%%%%%%%%%%%%%%%%%%%

Figure~\ref{fig:2O_FIG_02} highlights two primary effects of the combined analysis compared to the lepton-only analysis: the allowed blue regions slightly contract and shift relative to the red regions of the lepton-only case. The change in the preferred values of the lepton Lagrangian parameters is clearly visible in the one-dimensional projections for each column, with the most significant shift observed for $g\, v_u/ \sqrt{\Lambda}$. Additionally, the one-sigma regions from both analyses consistently overlap,
indicating full consistency between the results.

The situation is different for quarks, as shown in Figure~\ref{fig:2O_FIG_03}.
Here, the blue region from the combined analysis is considerably smaller than the green region corresponding to the quark-only analysis. Furthermore, the one-dimensional projections, particularly for  $g_2^d/g_1^d$ and $g_3^d/g_1^d$, reveal a significant shift in the preferred values from the fit.

\begin{figure}[h!]
\centering
\includegraphics[width=\linewidth]{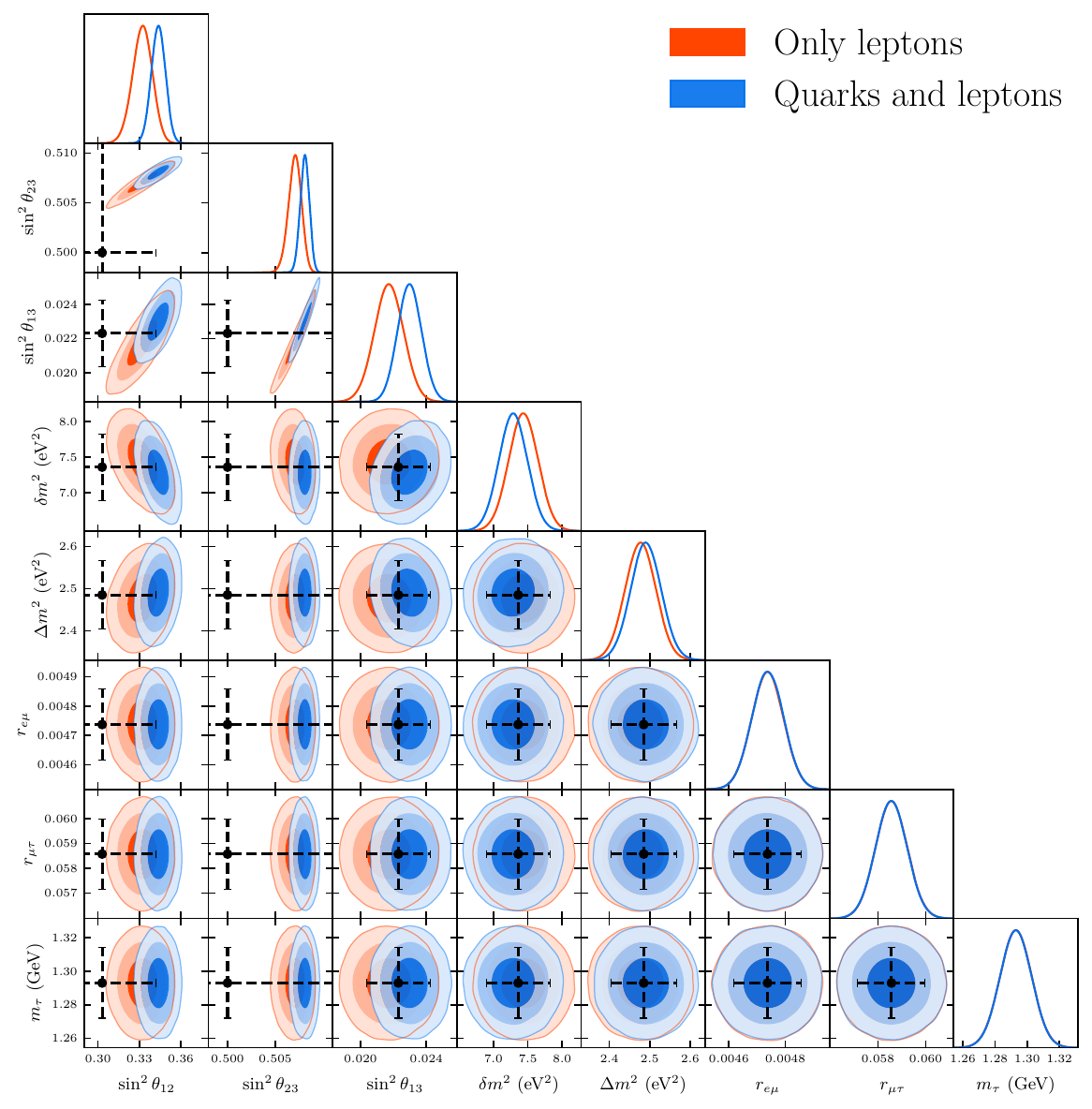}
\caption{Allowed regions for the lepton observables in the combined analyses. Different color shadings correspond to the 1$\sigma$, 2$\sigma$, and 3$\sigma$ confidence levels. The black dot indicates the current experimental measurement, and the dashed lines show the $3\sigma$ uncertainty range.}
\label{fig:2O_FIG_04}
\end{figure}

\begin{figure}[h!]
\centering
\includegraphics[width=\linewidth]{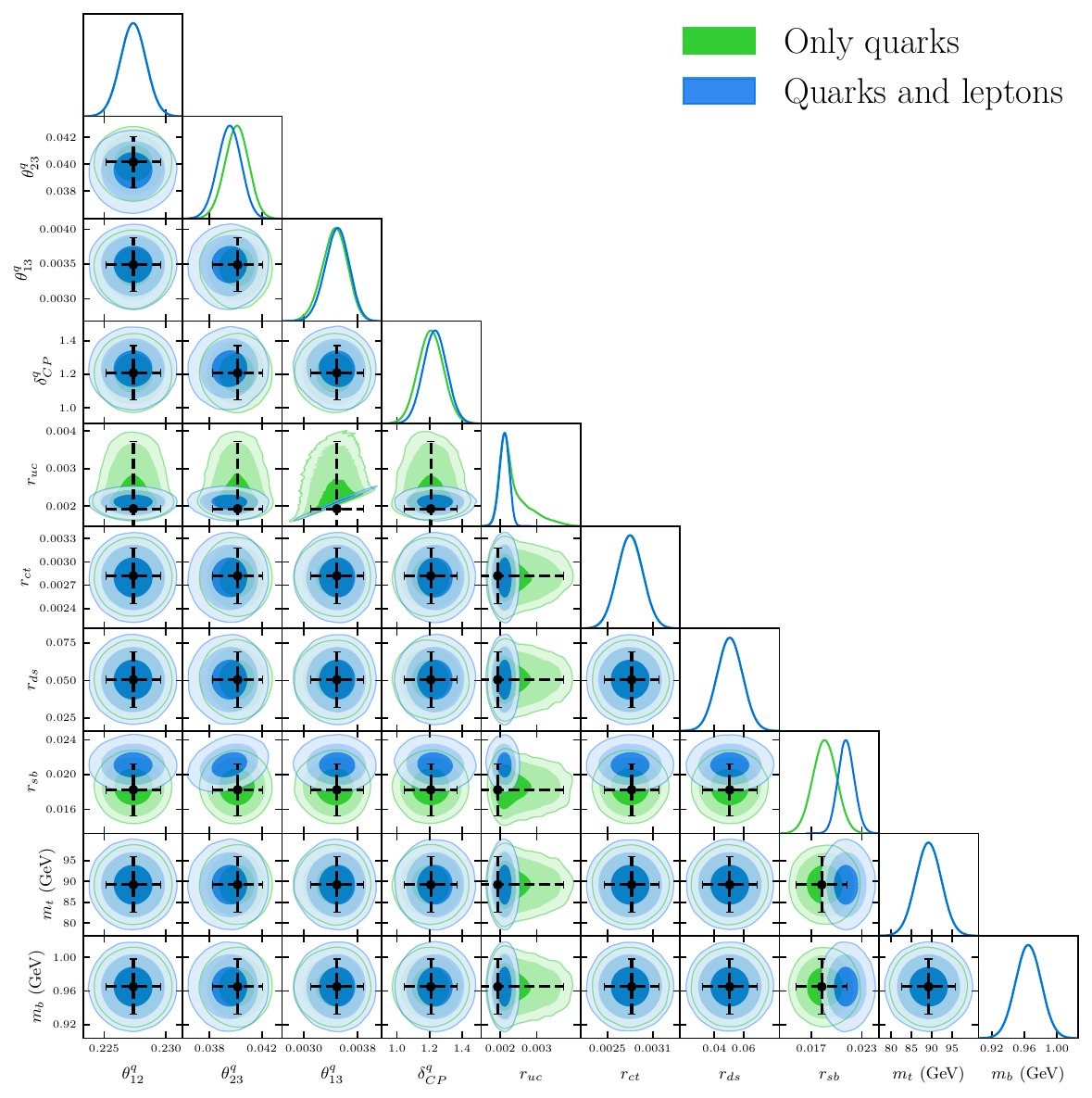}
\caption{Allowed regions for the quark observables in the combined analyses. Different color shadings correspond to the 1$\sigma$, 2$\sigma$, and 3$\sigma$ confidence levels. The black dot indicates the current experimental measurement, and the dashed lines show the $3\sigma$ uncertainty range.}
\label{fig:2O_FIG_05}
\end{figure}

Figures~\ref{fig:2O_FIG_04} and~\ref{fig:2O_FIG_05} show the allowed regions
for the lepton and quark observables, respectively. In Figure~\ref{fig:2O_FIG_04}, the combined fit is superimposed on the
lepton-only results. While $r_{e\mu}$, $r_{\mu\tau}$, $m_\tau$, and, to a slightly lesser extent, $\Delta m^2$ are well fitted in both the lepton-only and combined analyses, the fit quality decreases for the three mixing angles and $\delta m^2$, as clearly illustrated by their respective $\chi^2$ contributions in Table~\ref{tab:tab3}. In particular, the theoretical prediction for $\sin^2\theta_{12}$, which deviates by 2$\sigma$ from the experimental value in the lepton-only case, is found to be 3.2$\sigma$ off in the combined analysis. The model's prediction for $\sin^2\theta_{12}$ will be either confirmed or excluded by the JUNO experiment, which is
expected to measure it with sub-percent precision~\cite{JUNO:2022mxj,Capozzi:2015bpa}.

Figure~\ref{fig:2O_FIG_05} shows the allowed regions for the quark observables, with the combined fit superimposed on the quark-only results.
It is evident that all observables, except for $r_{uc}$, $r_{sb}$, $\delta^q_{CP}$, and $\theta_{23}^q$, are equally well fitted in both cases.
In particular, the preferred value of $r_{sb}$ in the combined fit is approximately $2.7\sigma$ away from the experimental value, whereas
it is perfectly accommodated by the quark-only analysis. The $\chi^2$ contribution for $r_{uc}$, $\delta^q_{CP}$, and $\theta_{23}^q$ increases slightly in the combined analysis, due to both a shift in the preferred theoretical value and the respective experimental uncertainties.

In summary, the results of the combined analysis clearly differ from the results of the lepton-only and quarks-only analyses. This is a consequence of  the existence of an interplay between the lepton and quark observables in the joint fit. The combined analysis accommodates all quark and lepton observables very well, producing theoretical values that closely match
the experimental measurements. However, due to the shift in the value of $\tau$ compared to the values found in the separate analyses, two observables, $\sin^2\theta_{12}$ and $r_{sb}$, are fitted slightly worse,
deviating from the experimental values by approximately $3.2\sigma$ and $2.7\sigma$, respectively.

\begin{figure}[h!]
\centering
\hspace*{-0.14\textwidth}  % Adjust this value to fine-tune the centering
\includegraphics[width=1.4\textwidth]{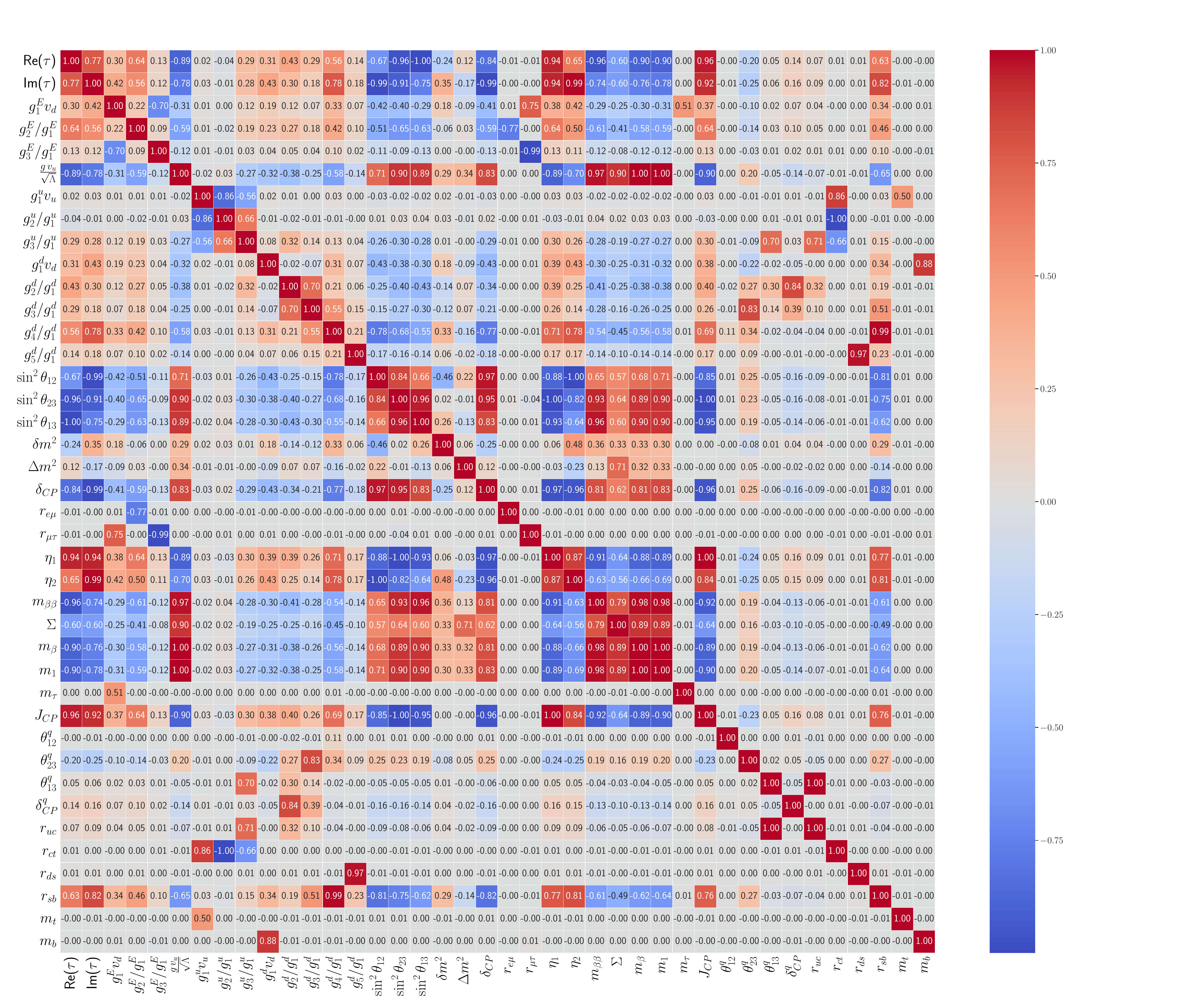}
\caption{Matrix of the correlations among parameters and observables.}
\label{fig:2O_FIG_06}
\end{figure}

The great advantage of the combined analysis is that it allows for a deeper
understanding of the fit by examining the correlations between the model
parameters and observables near the best-fit point, as shown in Figure~\ref{fig:2O_FIG_06}.

First of all, we observe that $r_{sb}$ is positively correlated with both
$\text{Re}(\tau)$ and $\text{Im}(\tau)$, whereas the opposite is true for
$\sin^2\theta_{12}$, which is negatively correlated with the real and
imaginary parts of $\tau$. When moving from the lepton-only analysis to the combined fit, both $\text{Re}(\tau)$ and $\text{Im}(\tau)$ decrease slightly.
As a result, $\sin^2\theta_{12}$ increases and deviates further from the
best-fit value due to its anticorrelation with $\text{Re}(\tau)$ and
$\text{Im}(\tau)$. Conversely, $r_{sb}$ also increases, in line with its positive correlation, but this also worsens the fit.

It is also possible to understand how the fit behaves by examining the
correlations between $\tau$ and the Lagrangian parameters, as well as the
correlations between these parameters and the observables. In particular, $r_{sb}$ is positively correlated with all quark model parameters. From Figure~\ref{fig:2O_FIG_03}, we can see that the combined fit increases
the values of $g_3^u/g_1^u$, $g_1^d v_d$, $g_2^d/g_1^d$, $g_3^d/g_1^d$, and
$g_4^d/g_1^d$, leading to an increase in $r_{sb}$ that worsens the fit.
In contrast, $\sin^2\theta_{12}$ is anticorrelated with $g^E_1 v_d$,
$g^E_2/g^E_1$, and $g^E_3/g^E_1$, but positively correlated with
$g v_u/\sqrt{\Lambda}$. From Figure~\ref{fig:2O_FIG_02}, we observe that
$g^E_1 v_d$, $g^E_2/g^E_1$, and $g^E_3/g^E_1$ decrease in the combined fit
compared to the lepton-only case, while $g v_u/\sqrt{\Lambda}$
increases. Considering these correlations with $\sin^2\theta_{12}$,
the changes in the lepton Lagrangian parameters in the combined fit result
in an increase in $\sin^2\theta_{12}$, thereby worsening the fit.

\begin{figure}[h!]
\centering
\includegraphics[width=\linewidth]{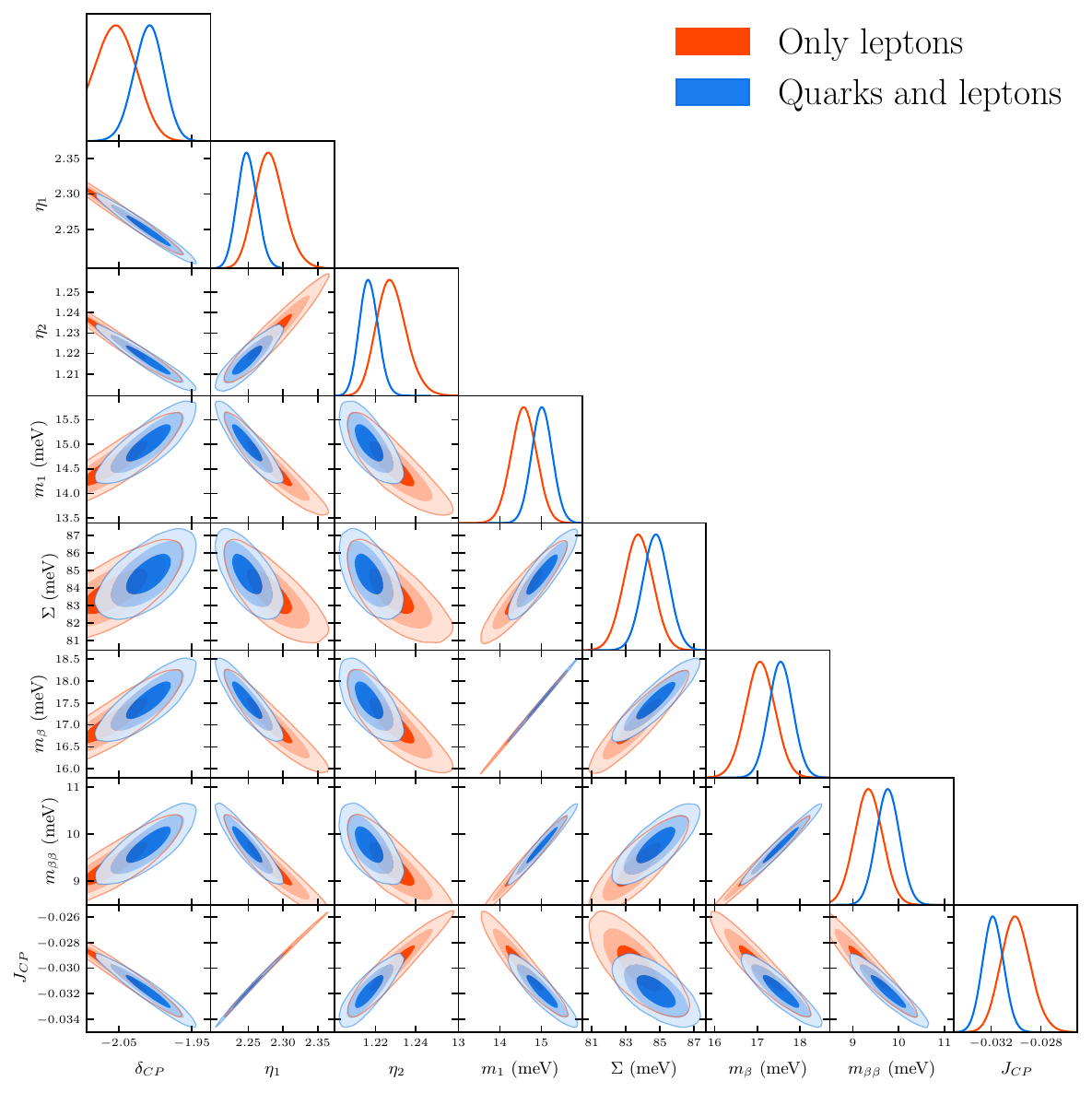}
\caption{Allowed regions for the unknown lepton observables. Different color shadings correspond to the 1$\sigma$, 2$\sigma$, and 3$\sigma$ confidence levels.}
\label{fig:2O_FIG_07}
\end{figure}

Figure~\ref{fig:2O_FIG_07} shows the allowed regions for the neutrino
observables that remain unknown, with the exception of $\delta_{CP}$, for which there are currently indications of a CP-violating value around $1.24\pi$ in Normal Ordering (and $1.52\pi$ in Inverted Ordering)~\cite{Capozzi:2021fjo}. The other observables shown in Figure~\ref{fig:2O_FIG_07} include the two Majorana phases ($\eta_1,\eta_2$)
\footnote{The parameterization we use for the Majorana phases, $\eta_1$ and $\eta_2$, as well as for the PMNS mixing matrix, strictly follows the conventions set by the PDG~\cite{PDG2024}.} ~\cite{Bilenky:1980cx}, the lightest neutrino mass $m_1$, the CPV rephasing invariant associated with the Dirac phase $\delta_{CP}$, $J_{CP} = c_{12} c_{13}^2 c_{23} s_{12} s_{13} s_{23}\sin{\delta_{CP}}$~\footnote{Together with a factor which depends on the neutrino mass squared differences, the energy and the distance travelled by the neutrinos, the $J_{CP}$ invariant controls the magnitude of the CP-violation effects in neutrino oscillations~\cite{Krastev:1988yu}. It is analogous to the CPV invariant in the quark sector introduced by Jarlskog~\cite{Jarlskog:1985cw}.}, and the three neutrino absolute mass observables
%%%%%%%%%%%%%%%%%%%%%%%%%%%%%%
\begin{align}
\Sigma &= m_1 + m_2 + m_3\ ,\\
 m_\beta &=
\left[c^2_{13}c^2_{12}m^2_1+c^2_{13}s^2_{12}m^2_2+s^2_{13}m^2_3
\right]^\frac{1}{2}\ ,\\
m_{\beta\beta} &= \left|
c^2_{13}c^2_{12}m_1+c^2_{13}s^2_{12}m_2e^{2i(\eta_2-\eta_1)}+s^2_{13}m_3
e^{-2i(\delta_{CP}+\eta_1)}\right|\ ,
\end{align}
%%%%%%%%%%%%%%%%%%%%%%
which are constrained by cosmology and astrophysics measurements, beta decay, and neutrinoless double-beta decay experiments, respectively. From the combined analysis, we get the best-fit values
%%%%%%%%%%%%%%%%%%%%%%%%%%%%%%%%%%%%%
\begin{align}
&\delta_{CP} = -2.01 \sim -0.64\pi\ ,\nonumber\\
&\eta_1 = 2.25 \sim 0.716\pi\ ,\nonumber\\
&\eta_2 = 1.22 \sim 0.387\pi\ ,\nonumber\\
&m_1 = 15.05 \text{ (meV)}\ ,\nonumber\\
&\Sigma =84.8 \text{ (meV)}\ ,\nonumber\\
&m_{\beta} = 17.6 \text{ (meV)}\ ,\nonumber\\
&m_{\beta\beta} = 9.75 \text{ (meV)}\ ,\nonumber\\
&J_{CP} = -0.032\ .
\label{eq:predict}
\end{align}
%%%%%%%%%%%%%%%%%%%%%%%%%%%%%%%
We observe from Figure~\ref{fig:2O_FIG_07} that, when moving from the lepton-only to the combined analysis, the $J_{CP}$ invariant and the two Majorana phases $\eta_1$ and $\eta_2$ slightly decrease, while all other observables show a slight increase in their best-fit values. This behavior can also be understood in terms of the correlations shown in Figure~\ref{fig:2O_FIG_07} between these unknowns and $\sin^2\theta_{12}$, the neutrino observable that changes the most in the combined analysis compared to the lepton-only case.

Moreover, it follows from Figure~\ref{fig:2O_FIG_07} that there are
\begin{enumerate}[label=\roman*),  itemsep=0pt, topsep=0pt, parsep=0pt, partopsep=0pt]
\item strong positive correlations between each pair of the three lepton (neutrino) mixing angles (this is also evident in Figure~\ref{fig:2O_FIG_04}),
\item strong positive (negative) correlations between each of the three lepton mixing angles and the leptonic Dirac (Majorana) CPV phase (phases) $\delta_{CP}$ ($\eta_1$, $\eta_2$),
\item strong positive correlations between each of the three lepton (neutrino) mixing angles and $m_1$ ($\Sigma$), $m_\beta$ and $m_{\beta\beta}$.
\end{enumerate}
Regarding the correlations between the quark and lepton observables,
the ratio $r_{sb}$ is
\begin{enumerate}[label=\roman*),  itemsep=0pt, topsep=0pt, parsep=0pt, partopsep=0pt]
\item strongly negatively correlated with each of the three lepton mixing angles,
\item strongly negatively (positively) correlated withthe leptonic
Dirac (Majorana) CPV phase (phases) $\delta_{CP}$ ($\eta_1$, $\eta_2$),
\item strongly negatively correlated with $m_1$ ($\Sigma$), $m_\beta$ and $m_{\beta\beta}$,
\item strongly positively correlated with the leptonic $J_{CP}$ factor.
 \end{enumerate}
A confirmation of the existence of correlations of the type listed above
would be a strong indication of the existence of an underlying modular flavour symmetry in particle physics.

Finally, note that the predicted range of $m_{\beta\beta}$ (9--11 meV) is still well below the current sensitivity of neutrinoless double beta decay experiments, which is around 30 meV for the most favorable nuclear matrix elements~\cite{Lisi:2022nka}. However, it is approaching the discovery potential of future ton-scale projects~\cite{Lisi:2023amm,guenette2024}.

\section[Conclusions]{Conclusions}\label{sec:conclusions}

In a predictive theory of flavour based on modular invariance - a promissing approach to the highly challenging and still unresolved
fundamental flavour problem in particle physics that has been actively developed in the last six years - there should exist an interplay and correlations between the quark and lepton observables. So far, however, the quark and lepton flavour observables have been analised separately in modular flavour models. This made it impossible to study the interconnection between these two sets of observables and also raises question about the validity of
the results of such  ``decoupled'' analyses, since a global description is rather nontrivial.

In the present work we have presented the first combined analysis of quark and lepton observables  in flavour models based on modular invariance.
The analysis is performed within a specific model with modular (flavour) $2O$ symmetry, which is unique in its class by describing the twenty-two observables in the quark and lepton sectors (eighteen of which have been measured) by altogether only fourteen real parameters - the minimal number of real parameters in the flavour models based on modular invariance constructed so far. The modular (flavour) and CP symmetries are broken in the model by the vacuum expectation value (VEV) of a single complex scalar field - the modulus $\tau$.

In order to assess the interplay and correlations between the quark and lepton observables, we performed first separate analyses of the lepton sector and of the quark sector, which was followed by a joint analysis of both
the quark and lepton sectors.  We have shown that the results of the
combined analysis clearly differ at the level of detailed best-fit values and jointly allowed regions from the results of the lepton-only and quarks-only analyses.
This is a consequence of the existence of an
interplay between the lepton and quark observables in the joint fit. We have discussed the indicated differences in detail.

More specifically, we have shown that the minimum of the combined $\chi^2_\text{comb}$ is not merely the sum of the $\chi^2$ minima obtained in the lepton-only and quark-only analyses (approximately 4.74), but is significantly higher in the combined case (Table~\ref{tab:tab3}). This suggests a complex interplay between the lepton and quark sectors, as their respective minima occupy distinct regions of the $\tau$ plane. Consequently, the combined analysis identified a new global minimum for $\tau$, which minimizes the total $\chi^2$ contribution for all eighteen observables, resulting in the overall $\chi^2_\text{comb}$. We have shown further that with respect to the results of the lepton-only analysis, the fit quality decreases for the three neutrino mixing angles and $\delta m^2$ in the joint lepton and quark analysis (Table~\ref{tab:tab3}). In particular, the theoretical prediction for $\sin^2\theta_{12}$, which deviates by $\sim 2\sigma$ from the experimental value in the lepton-only analysis case, is found to be $\sim 3.2\sigma$ off in the combined analysis. We found similar results for the quark observables $r_{uc}$, $r_{sb}$ - the ratios of  up and charm qaurk masses and of strange and bottom quark masses, $\delta^q_{CP}$ and $\theta_{23}^q$. In particular, the preferred value of $r_{sb}$ in the combined fit is approximately $2.7\sigma$ away from the experimental value, whereas it is perfectly accommodated by the quark-only analysis.

In addition the results of our joint analysis (presented graphically in Figure~\ref{fig:2O_FIG_07}) showed that there are
\begin{enumerate}[label=\roman*),  itemsep=0pt, topsep=0pt, parsep=0pt, partopsep=0pt]
\item strong positive correlations between each pair of the three lepton (neutrino) mixing angles (this is also evident in Figure~\ref{fig:2O_FIG_04}),
\item strong positive (negative) correlations between each of the three lepton mixing angles and the leptonic Dirac (Majorana) CPV phase (phases) $\delta_{CP}$ ($\eta_1$, $\eta_2$),
\item strong positive correlations between each of the three lepton (neutrino) mixing angles and $m_1$ ($\Sigma$), $m_\beta$ and $m_{\beta\beta}$.
\end{enumerate}

Regarding the correlations between the quark and lepton observables,
we found, in particular, that the ratio $r_{sb}$ is
\begin{enumerate}[label=\roman*),  itemsep=0pt, topsep=0pt, parsep=0pt, partopsep=0pt]
\item strongly negatively correlated with each of the three lepton mixing angles,
\item strongly negatively (positively) correlated withthe leptonic Dirac (Majorana) CPV phase (phases) $\delta_{CP}$ ($\eta_1$, $\eta_2$),
\item strongly negatively correlated with $m_1$ ($\Sigma$), $m_\beta$ and $m_{\beta\beta}$,
\item strongly positively correlated with the leptonic $J_{CP}$ factor.
 \end{enumerate}
A confirmation of the existence of correlations of the type listed above (e.g., through more precise or new experimental data aligned along the main correlation directions) would be a strong indication in favour of the examined model and of its underlying modular symmetry.

It follows from the results of the joint lepton and quark observables analysis that overall the model not only relatively accurately reproduces the established observables, such as quark masses, CKM mixing angles, and
lepton flavor data, but also offers precise predictions for quantities
that are yet to be experimentally determined, including the phase $\delta_{CP}$, the Majorana phases, the absolute neutrino mass $m_1$, and the effective masses $m_\beta$ and $m_{\beta\beta}$, which are of significance in the context of beta decay and neutrinoless double-beta decay experiments.
These predictions fall within ranges that are likely to be probed by ongoing and future experimental efforts.

While the overall agreement between the model and current data is excellent,
certain tensions remain, particularly in the case of $\sin^2\theta_{12}$ and
$r_{sb}$, which exhibit deviations of approximately $3.2\sigma$ and
$2.7\sigma$, respectively, from their current experimental counterparts. These tensions may be tested by future measurements or possibly alleviated in  model variants. Nonetheless, the model's ability to correlate the quark and lepton sectors through the modular parameter $\tau$ provides a compelling and economical approach to the flavor problem, reinforcing the potential of modular symmetries as a powerful tool in particle physics. Upcoming neutrino oscillation experiments, such as JUNO, will be instrumental in testing the model predictions, particularly in relation to $\sin^2\theta_{12}$, which JUNO will measure with unprecedented precision.

Thus, the results of our study showed that in modular flavour models there exist a strong interplay and correlations between the quark and lepton observables, that cannot be appreciated in separate analyses of the two flavor sectors.	Not taking them into account through a joint analysis
may lead to incorrect conclusions about the viability of the tested model.

Overall, our study also underscores the effectiveness of modular symmetries in offering a unified description of quark and lepton flavors, while also laying the foundation for future theoretical developments in the
fields of neutrino and flavour physics.

\subsection*{Acknowledgements}
The work of S.T.P.\ was supported in part by the European Union's Horizon 2020 research and innovation programme under the Marie Sk\l{}odowska-Curie grant agreement No.~860881-HIDDeN, by the Italian INFN program on Theoretical Astroparticle Physics and by the World Premier International Research Center Initiative (WPI Initiative, MEXT), Japan. The work of E.L.\ and A.M.\ was partially supported by the research grant number 2022E2J4RK ``PANTHEON: Perspectives in Astroparticle and Neutrino THEory with Old and New messengers'' under the program PRIN 2022 funded by the Italian Ministero dell'Universit\`a e della Ricerca (MUR) and by the European Union -- Next Generation EU, as well as by the Theoretical Astroparticle
Physics (TAsP) initiative of the Istituto Nazionale di Fisica Nucleare (INFN). G.J.D.\ is supported by the National Natural Science Foundation of China under Grant No.~2375104.

\clearpage


\clearpage

\section*{Appendix}

\setcounter{equation}{0}
\renewcommand{\theequation}{\thesection.\arabic{equation}}

\begin{appendix}

\section{\label{app:Modular-forms} Vector valued modular form multiplets of $2O$}

The group $2O$ can be generated by the modular generators $S$ and $T$ obeying the relations~\cite{Liu:2021gwa,Ding:2023ydy}:
%\begin{equation}
%\label{eq:2Ogroup}
%S^2=T^4=R,~~~ RT=TR,~~~R^2=(ST)^3=1\,,
%\end{equation}
%or equivalently
\begin{equation}
\label{eq:2Ogroup}
S^4=(ST)^3=S^2T^4=1,\quad S^2T=TS^2\,.
\end{equation}
The finite modular group $2O$ has eight irreducible representations: two singlets $\bm{1}$, $\bm{1}'$, three doublets $\bm{2}$, $\widehat{\bm{2}},\widehat{\bm{2}}'$, two triplets $\bm{3}$ and $\bm{3}'$, and a quartet $\widehat{\bm{4}}$. The representation matrices of the generators $S$ and $T$ in different representations are given by
\begin{eqnarray}
\nonumber\mathbf{1}: ~& S=1, ~&~ T=1\,, \\
\nonumber\mathbf{1}':~& S=-1, ~&~ T=-1\,, \\
\nonumber\mathbf{2}: ~&~ S=-\frac{1}{2}
\begin{pmatrix}
1  ~&~  \sqrt{3}  \\
\sqrt{3}  ~&~  -1
\end{pmatrix}, & T=\begin{pmatrix}
1 ~&~ 0 \\
0 ~&~ -1 \\
\end{pmatrix} \,,\\
\nonumber\mathbf{\widehat{2}}: ~&~ S=\frac{i}{\sqrt{2}}
\begin{pmatrix}
1  ~&~  1  \\
1  ~&~  -1
\end{pmatrix}, & T=\begin{pmatrix}
\xi^{3} ~&~ 0 \\
0 ~&~ \xi^{5} \\
\end{pmatrix}\,,\\
\nonumber \mathbf{\widehat{2}}': ~&~ S=\frac{i}{\sqrt{2}}
\begin{pmatrix}
1  ~&~  1  \\
1  ~&~  -1
\end{pmatrix}, & T=\begin{pmatrix}
\xi ~&~ 0 \\
0 ~&~ \xi^{7} \\
\end{pmatrix} \,,\\
\nonumber\mathbf{3}: ~& S=\frac{1}{2}\begin{pmatrix}
0~&\sqrt{2}~&\sqrt{2}\\
\sqrt{2}~&-1~&1\\
\sqrt{2}~&1~&-1
\end{pmatrix},
 ~&~ T=\begin{pmatrix}
1~&0~&0\\
0~&\xi^{2}~&0\\
0~&0~&\xi^6
\end{pmatrix} \,,\\
\nonumber \mathbf{3'}: ~& S=-\frac{1}{2}\begin{pmatrix}
0~&\sqrt{2}~&\sqrt{2}\\
\sqrt{2}~&-1~&1\\
\sqrt{2}~&1~&-1
\end{pmatrix},
 ~&~ T=-\begin{pmatrix}
1~&0~&0\\
0~&\xi^{2}~&0\\
0~&0~&\xi^6
\end{pmatrix} \,,\\
\label{eq:irr-reps} \mathbf{\widehat{4}}: ~& S=\dfrac{i}{2 \sqrt{2}}
\begin{pmatrix}
1 ~& 1 ~& \sqrt{3} ~& -\sqrt{3} \\
1 ~& -1 ~& -\sqrt{3} ~& -\sqrt{3} \\
\sqrt{3} ~& -\sqrt{3} ~& 1 ~& 1 \\
-\sqrt{3} ~& -\sqrt{3} ~& 1 ~& -1 \\
\end{pmatrix},
 ~&~ T=\begin{pmatrix}
\xi^{5}~&0~&0&~0\\
0~&\xi^{3}~&0&~0\\
0~&0~&\xi^7&~0\\
0~&0~&0&~\xi\\
\end{pmatrix} \,,~~~
\end{eqnarray}
where $\xi=e^{i\pi/4}$. Notice the representation matrix $\rho(R)=1$ in the unhatted irreps $\bm{1}$, $\bm{1}'$, $\bm{2}$, $\bm{3}, \bm{3}'$ and $\rho(R)=-1$ in the hatted irreps $\widehat{\bm{2}}$, $\widehat{\bm{2}}'$, $\widehat{\bm{4}}$. In the following, we report the $q$-expansion of modular form multiplets of $2O$.

\begin{itemize}

\item{$k=2$}

There is a doublet modular form $Y^{(2)}_{\mathbf{2}}(\tau)$ and a triplet modular form $Y^{(2)}_{\mathbf{3}}(\tau)$ at weight $2$,
\begin{eqnarray}
\nonumber Y^{(2)}_{\mathbf{2}}(\tau)&=&\begin{pmatrix}
 1 + 24 q + 24 q^2 + 96 q^3 + 24 q^4 + 144 q^5 + 96 q^6+\dots \\
 8 \sqrt{3} q^{1/2} \left(1 + 4 q + 6 q^2 + 8 q^3 + 13 q^4 + 12 q^5 + 14 q^6 +\dots \right)\\
\end{pmatrix}\,,\\
\label{eq:MF-k2} Y^{(2)}_{\mathbf{3}}(\tau)&=&\begin{pmatrix}
 1 - 8 q + 24 q^2 - 32 q^3 + 24 q^4 - 48 q^5 + 96 q^6+\dots\\
-4 \sqrt{2} q^{1/4} \left(1 + 6 q + 13 q^2 + 14 q^3 + 18 q^4 + 32 q^5 + 31 q^6 +\dots\right)\\
-16 \sqrt{2} q^{3/4} \left(1 + 2 q + 3 q^2 + 6 q^3 + 5 q^4 + 6 q^5 + 10 q^6 +\dots \right)
\end{pmatrix}\,,~~~~~~~~
\end{eqnarray}
with $q=^{2\pi i\tau}$.

\item{$k=3$}

There is only a quartet modular form $Y^{(3)}_{\widehat{\mathbf{4}}}(\tau)$ at weight 3,
\begin{equation}
\label{eq:MF-k3} Y^{(3)}_{\widehat{\mathbf{4}}}(\tau)=\begin{pmatrix}
4\sqrt{3}\,q^{5/8} \left(1-q-4 q^2+3 q^3+q^4+3 q^5+13q^6+\dots\right) \\
2\sqrt{3}\,q^{3/8} \left(1+q-7 q^2-6 q^3+16q^4+9 q^5-6q^6\dots\right) \\
-8 q^{7/8} \left(1-3q+ 3q^2-4q^3+3q^4+6q^5-3q^6+\dots\right) \\
q^{1/8} \left(1+3q-6q^2-23q^3+12q^4+66q^5-15q^6+\dots\right)
\end{pmatrix}\,.
\end{equation}

\item{$k=4$}

\begin{eqnarray}
\nonumber Y^{(4)}_{\mathbf{1}}(\tau)&=&1+240 q+2160 q^2+6720 q^3+17520 q^4+30240 q^5+60480 q^6+\ldots\,,\\
\nonumber Y^{(4)}_{\mathbf{2}}(\tau)&=&\begin{pmatrix}
-1+144 q+912 q^2+4032 q^3+7056 q^4+18144 q^5+25536 q^6+\ldots \\
 16 \sqrt{3}\, q^{1/2} \left(1+28 q+126 q^2+344 q^3+757 q^4+1332 q^5+2198 q^6+\ldots\right)
\end{pmatrix}\,,\\
\nonumber Y^{(4)}_{\mathbf{3}}(\tau)&=&\begin{pmatrix}
1+16 q-144 q^2+448 q^3-912 q^4+2016 q^5-4032 q^6+\ldots \\
2\sqrt{2}\,q^{1/4}\left(1+126q+757 q^2+2198 q^3+4914 q^4+9632 q^5+15751 q^6+\ldots\right) \\
8 \sqrt{2}\,q^{3/4} \left(7+86 q+333 q^2+882 q^3+1715 q^4+3042 q^5+5110 q^6+\ldots\right)
\end{pmatrix}\,,\\
Y^{(4)}_{\mathbf{3'}}(\tau)&=&\begin{pmatrix}
2 \sqrt{2}\,q^{1/2} \left(1 - 4 q - 2 q^2 + 24 q^3 - 11 q^4 - 44 q^5 + 22 q^6+\dots\right) \\
 4 q^{3/4} \,\left(1 - 6 q + 11 q^2 - 2 q^3 - 11 q^4 + 14 q^5 - 38 q^6 +\dots\right) \\
 q^{1/4}\,\left(-1 + 2 q + 11 q^2 - 22 q^3 - 50 q^4 + 96 q^5 + 121 q^6 +\dots\right)
\end{pmatrix}\,.
\end{eqnarray}

\item{$k=5$}

\begin{eqnarray}
\nonumber Y^{(5)}_{\widehat{\mathbf{2}}}(\tau)&=&\begin{pmatrix}
 q^{3/8} \left(1 - 7 q + 9 q^2 + 42 q^3 - 112 q^4 - 63 q^5 + 378 q^6 + \dots\right) \\
 2 q^{5/8} \left(1 - 9 q + 28 q^2 - 21 q^3 - 63 q^4 + 123 q^5 - 35 q^6+\dots\right)\\
\end{pmatrix}\,,\\
\nonumber Y^{(5)}_{\widehat{\mathbf{2}}'}(\tau)&=&\begin{pmatrix}
q^{1/8} \left(1 + 123 q + 378 q^2 - 191 q^3 - 1428 q^4 - 1134 q^5 - 735 q^6+\dots\right) \\
8 q^{7/8} \left(7 + 51 q - 27 q^2 - 28 q^3 - 459 q^4 + 378 q^5 - 357 q^6+\dots\right)\\
\end{pmatrix}\,,\\
Y^{(5)}_{\widehat{\mathbf{4}}}(\tau)&=&\begin{pmatrix}
4 \sqrt{3}\,q^{5/8} \left(-3-37 q-20 q^2+63 q^3+253 q^4+207 q^5-471 q^6+\ldots\right) \\
-2\sqrt{3}\,q^{3/8}\left(1+57q+73 q^2-150 q^3-240 q^4-447 q^5-198 q^6+\ldots\right) \\
8q^{7/8}\left(5+9 q+63 q^2-212 q^3+111 q^4-306 q^5+513 q^6+\ldots\right) \\
q^{1/8}\left(1-69 q-198 q^2+193 q^3+684 q^4+18 q^5+801 q^6+\ldots\right)
\end{pmatrix}\,.\qquad
\end{eqnarray}

\item{$k=6$}

\begin{small}
\begin{eqnarray}
\nonumber Y^{(6)}_{\mathbf{1}}(\tau)&=&1-504q\left(1+33q+244q^2+1057q^3+3126q^4+8052q^5+\ldots\right)\,,\\
\nonumber Y^{(6)}_{\mathbf{1'}}(\tau)&=&q^{1/2} \left(1-12 q+54 q^2-88 q^3-99 q^4+540 q^5-418 q^6+\ldots\right)\,,\\
\nonumber Y^{(6)}_{\mathbf{2}}(\tau)&=&\begin{pmatrix}
1+264 q+7944 q^2+64416 q^3+253704 q^4+825264 q^5+1938336 q^6+\ldots \\
8 \sqrt{3} q^{1/2} \left(1+244 q+3126 q^2+16808 q^3+59293 q^4+161052 q^5+371294 q^6+\ldots\right)\\
\end{pmatrix}\,,\\
\nonumber Y^{(6)}_{\mathbf{3}I}(\tau)&=&\begin{pmatrix}
1+232 q+264 q^2-4832 q^3+7944 q^4-12048 q^5+64416 q^6+\ldots \\
-4\sqrt{2}\,q^{1/4} \left(1+246 q+3613 q^2+22814 q^3+89298 q^4+257312 q^5+610351 q^6+\ldots\right) \\
-16\sqrt{2}\, q^{3/4}\left(1+242 q+2643 q^2+11766 q^3+38885 q^4+99606 q^5+226090 q^6+\ldots\right)
\end{pmatrix}\,,\\
\nonumber Y^{(6)}_{\mathbf{3}II}(\tau)&=&\begin{pmatrix}
\sqrt{2}\left(-1+56 q-264 q^2+1376 q^3-7944 q^4+27600 q^5-64416 q^6+\ldots\right) \\
q^{1/4} \left(-1+1482 q+29795 q^2+186274 q^3+709038 q^4+2048992 q^5+4884689 q^6+\ldots\right) \\
4q^{3/4}\left(35+2134 q+19929 q^2+95586 q^3+309199 q^4+806082 q^5+1799486 q^6+\ldots\right)
\end{pmatrix}\,,\\
Y^{(6)}_{\mathbf{3'}}(\tau)&=&\begin{pmatrix}
4\sqrt{2}\,q^{1/2} \left(1+20 q-74 q^2-24 q^3+157 q^4+124 q^5+478 q^6+\ldots\right) \\
4q^{3/4}\left(5-6 q+31 q^2-370 q^3+761 q^4+46 q^5-430 q^6+\dots\right) \\
q^{1/4} \left(1-74 q+157 q^2+478 q^3-1198 q^4-480 q^5+2351 q^6+\ldots\right)
\end{pmatrix}\,.
\end{eqnarray}
\end{small}

\end{itemize}

The normalisation of the modular forms
\footnote{A detailed discussion of the problem of normalisation of
the modular forms is presented in~\cite{Petcov:2023fwh}.} we are employing
is determined by their series expansions given above. More specifically,
the multiplets of modular forms are normalised in such a way that in the limit of large $\text{Im}(\tau)$
%imaginary part $y$, at leading order each of them
the largest component is proportional (up to a  possible order one constant)
to $q^r$ at leading order, where $r$ is some rational number or zero.
For example,
%%%%%%%%%%%%%%%%%%%%%%%%
\begin{eqnarray}
\nonumber
Y^{(2)}_2(\tau) &\approx& (1, 0)\,,\\
\nonumber
Y^{(2)}_3(\tau) &\approx& (1, 0, 0)\,,\\
Y^{(3)}_{\widehat{4}}(\tau) &\approx& (0, 0, 0, q^{1/8})\,,
\label{eq:norm}
\end{eqnarray}
for $\text{Im}(\tau)\gg1$.
%%%%%%%%%%%%%%%%%%%%%%%%
%
The normalisation of the lowest weight 2 modular forms, in particular,
coincides with the normalisation used in~\cite{Novichkov:2020eep}
but is rescaled by the universal constant factor $1/\sqrt{2}$.

\end{appendix}

%\bibliographystyle{utphys}
%\bibliography{references}
%\end{document}

\providecommand{\href}[2]{#2}\begingroup\raggedright\endgroup

\end{document}